\documentclass[a4paper,11pt]{article}
\usepackage[utf8]{inputenc}
\usepackage[margin=2.5cm]{geometry}
\usepackage[style=nature]{biblatex}
\usepackage{xcolor}
\usepackage{graphicx}
\usepackage{amsmath,amssymb}
\usepackage{amsfonts}
\usepackage[nice]{nicefrac}
\usepackage[title]{appendix}
\usepackage{hyperref}
\usepackage{authblk}
\usepackage[misc]{ifsym}
\usepackage{multirow}
\usepackage{booktabs}
\usepackage{lineno}

\title{\Huge Maximum size and magnitude of injection-induced slow slip events}

\author[$a$]{Alexis S\'aez$^1$\footnote{Present address: Division of Geological and Planetary Sciences, California Institute of Technology, Pasadena, California 91125, USA}}
\author[$b$]{François Passelègue}
\author[$a$]{Brice Lecampion}
\affil[$a$]{Ecole Polytechnique F\'ed\'erale de Lausanne (EPFL), Institute of Civil Engineering, Gaznat chair on Geo-Energy, CH-1015 Lausanne, Switzerland}
\affil[$b$]{Géoazur, Université Côte d'Azur, CNRS, Observatoire de la Côte d'Azur, IRD, Sophia Antipolis, Nice, France\\
\begin{center}
    $^1$E-mail: \href{mailto:alexis.saez@epfl.ch}{saez@caltech.edu}
\end{center}}

\date{}

\begin{document}

\setlength{\parskip}{5pt}

\maketitle


\begin{abstract}
    Fluid injections can induce aseismic slip, resulting in stress changes that may propagate faster than pore pressure diffusion, potentially triggering seismicity at significant distances from injection wells. Constraining the maximum extent of these aseismic ruptures is thus important for better delineating the influence zone of injections concerning their seismic hazard. Here we derive a scaling relation based on rupture physics for the maximum size of aseismic ruptures, accounting for fluid injections with arbitrary flow rate histories. Moreover, based on mounting evidence that the moment release during these operations is often predominantly aseismic, we derive a scaling relation for the maximum magnitude of aseismic slip events. Our theoretical predictions are consistent with observations over a broad spectrum of event sizes, from laboratory to real-world cases, indicating that fault-zone storativity, background stress state, and injected fluid volume are key determinants of the maximum size and magnitude of injection-induced slow slip events.
\end{abstract}

\section{Introduction}\label{sec:introduction}

A growing body of observations suggests that a significant part of the deformation induced by subsurface fluid injections is due to aseismic fault motions \cite{Hamilton_Meehan_1971,Scotti_Cornet_1994,Cornet_Helm_1997,Bourouis_Bernard_2007,Calo_Dorbath_2011,Guglielmi_Cappa_2015,Wei_Avouac_2015,Cappa_Scuderi_2019,Materna_Barbour_2022,Eyre_Samsonov_2022,Pepin_Ellsworth_2022}. This phenomenon, known as injection-induced aseismic slip, has been known since at least the 1960s when a slow surface fault rupture was causally linked to fluid injection operations of an oil field in Los Angeles \cite{Hamilton_Meehan_1971}. Since then, an increasing number of observational studies have inferred the occurrence of slow slip events as a result of industrial fluid injections. For example, in the Brawley geothermal field, California, ground- and satellite-based geodetic techniques allowed for the detection of an injection-induced aseismic slip event \cite{Wei_Avouac_2015,Materna_Barbour_2022}. This event was found to precede and likely trigger a seismic sequence in 2012 \cite{Im_Avouac_2021}. In western Canada, two of the largest aseismic slip events observed thus far (magnitudes 5.0 and 4.2) occurred in 2017-2018 and were detected using InSAR measurements of surface deformation \cite{Eyre_Samsonov_2022}. These events were attributed to hydraulic fractures possibly intersecting glide planes during the stimulation of an unconventional hydrocarbon reservoir \cite{Eyre_Samsonov_2022}. Similarly, InSAR-derived surface deformations allowed for the recent detection of aseismic ruptures in the southern Delaware Basin, Texas \cite{Pepin_Ellsworth_2022}, likely induced by wastewater injection operations \cite{Dvory_Yang_2022}. These recent geodetic observations, in combination with mounting evidence for aseismic slip from fluid-injection field experiments \cite{Scotti_Cornet_1994,Cornet_Helm_1997,Bourouis_Bernard_2007,Calo_Dorbath_2011,Guglielmi_Cappa_2015,Cappa_Scuderi_2019}, suggest that injection-induced slow slip events might be a ubiquitous phenomenon, largely underdetected over the past decades only due to the lack of geodetic monitoring.

There is increasing recognition of the importance of injection-induced aseismic slip in the geo-energy industry. For instance, in the development of deep geothermal reservoirs, hydraulic stimulation techniques are commonly used to reactivate pre-existing fractures in shear. This process aims to enhance reservoir permeability through the permanent dilation of pre-existing fractures or the creation of new ones. The occurrence of predominantly aseismic rather than seismic slip is desirable, as earthquakes of significant magnitude can pose a substantial hazard to the success of these projects \cite{Deichmann_Giardini_2009,Ellsworth_Giardini_2019}. Injection-induced aseismic slip can be, however,  detrimental in several ways. For example, aseismic slip on fractures intersecting wells can cause casing shearing \cite{Cornet_Helm_1997,Eyre_Samsonov_2022} and adversely impact well stability \cite{Dusseault_Bruno_2001,Li_Liu_2019}. Additionally, in CO$_2$ storage operations, injection-induced aseismic slip could affect the integrity of low--permeability caprocks, as fault slip may be accompanied by permeability enhancements, increasing the risk of CO$_2$ leakage \cite{Cappa_Rutqvist_2011,Rinaldi_Vilarrasa_2015}. Similar concerns may arise in other underground operations such as the storage of hydrogen and gas. Furthermore, it is well-established that quasi-static stress changes due to aseismic slip may induce seismic failures on nearby unstable fault patches \cite{Scotti_Cornet_1994,Cornet_Helm_1997,Bourouis_Bernard_2007,Wei_Avouac_2015,Guglielmi_Cappa_2015}. Moreover, since aseismic slip can propagate faster than pore pressure diffusion \cite{Guglielmi_Cappa_2015,Bhattacharya_Viesca_2019,Eyre_Eaton_2019}, aseismic-slip stress changes can potentially reach regions much further than the zones affected by the direct increase in pore pressure due to injection, thereby increasing the likelihood of triggering earthquakes of undesirably large magnitude by perturbing a larger rock volume \cite{Eyre_Eaton_2019,Im_Avouac_2021,Vilarrasa_DeSimone_2021}.

Understanding the physical factors controlling the spatial extent of aseismic slip is thus of great importance to better constrain the influence zone of injection operations concerning seismic hazards. Recent theoretical and numerical modeling studies have provided, within certain simplifying assumptions, a fundamental mechanistic understanding of how injection-induced aseismic slip grows in a realistic three-dimensional context and through all its stages, from nucleation to arrest \cite{Saez_Lecampion_2022,Saez_Lecampion_2023,Saez_Lecampion_2024}. 
Estimating the rupture run-out distance of aseismic slip transients remains, despite these efforts, an unresolved issue, particularly as these prior investigations have focused only on specific injection protocols \cite{Saez_Lecampion_2022,Saez_Lecampion_2023,Saez_Lecampion_2024}. Yet the spatiotemporal patterns of injection-induced aseismic slip growth are anticipated to be strongly influenced by the history of injection flow rate 
\cite{Saez_Lecampion_2022}. On the other hand, a related issue is estimating the maximum magnitude of injection-induced earthquakes. This quantity plays a crucial role in earthquake hazard assessment and has been the focus of significant research efforts in recent times \cite{Shapiro_Dinske_2007,Shapiro_Kruger_2011,McGarr_2014,Atkinson_Eaton_2016,vanderElst_Page_2016,McGarr_Barbour_2017,Galis_Ampuero_2017,DeBarros_Cappa_2019,Bentz_Kwiatek_2020,Li_Elsworth_2021,Shapiro_Kim_2021}. 
A common limitation of prior research in this area is neglecting the portion of moment release due to aseismic slip, despite substantial evidence suggesting that aseismic motions may contribute significantly to the total moment release \cite{Hamilton_Meehan_1971,Scotti_Cornet_1994,Cornet_Helm_1997,Bourouis_Bernard_2007,Calo_Dorbath_2011,Guglielmi_Cappa_2015,Wei_Avouac_2015,Cappa_Scuderi_2019,Materna_Barbour_2022,Eyre_Samsonov_2022,Pepin_Ellsworth_2022}, potentially surpassing seismic contributions in some cases \cite{Hamilton_Meehan_1971,Scotti_Cornet_1994,Cornet_Helm_1997,Bourouis_Bernard_2007,Calo_Dorbath_2011,Guglielmi_Cappa_2015,Cappa_Scuderi_2019,Im_Avouac_2021b,Eyre_Samsonov_2022}. Understanding the factors governing aseismic moment release is thus important, and would constitute a first step toward understanding the physical controls on slip partitioning, that is, the relative contributions of aseismic and seismic motions to the release of elastic strain energy, which is crucial for a better understanding of the seismic hazard posed by these operations.

Building upon our previous works \cite{Saez_Lecampion_2022,Saez_Lecampion_2023,Saez_Lecampion_2024}, we develop here an upper-bound model for the spatial extent and moment release of injection-induced aseismic slip events. Our model notably accounts for fluid injections that are conducted with an arbitrary history of injection flow rate including the shut-in stage, thus effectively reproducing aseismic slip events during their entire life cycle, from nucleation to arrest. Using fracture mechanics theory, scaling analysis, and numerical simulations, we propose scaling relations for the maximum size and magnitude of aseismic ruptures which are shown to be consistent with a global compilation of events that vary in size from cm-scale slip transients monitored in the laboratory to km-scale, geodetically inferred slow slip events induced by industrial injections. Our results suggest that fault-zone hydro-mechanical storativity, background stress state, and injected fluid volume are crucial quantities in determining upper limits for the size and magnitude of aseismic ruptures. Moreover, the total fluid volume injected by a given operation is shown to be the only operational parameter that matters in determining the upper bounds in our model, regardless of any other characteristic of the injection protocol.   

\section{Results}\label{sec:results}

\subsection{Physical model and upper bound rationale}\label{sec:model}

\begin{figure}
    \centering
    \includegraphics[width=15.5cm]{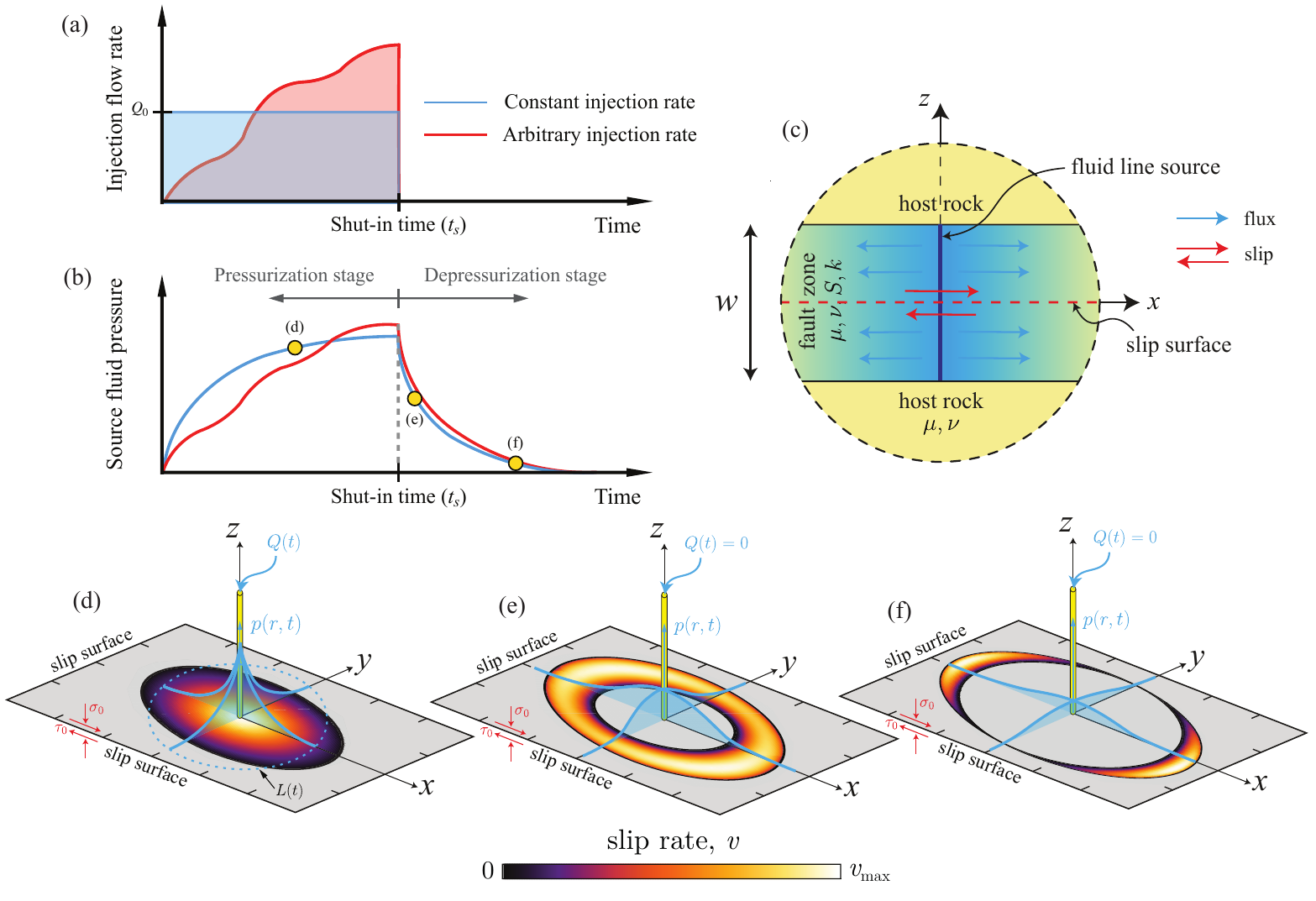}
    \caption{\textbf{Model schematics.} (a) Fluid is injected at a constant or arbitrary volumetric rate until the shut-in time $t_s$ at which the injection is instantaneously stopped.
    (b) This results in two distinct stages: a pressurization stage and a depressurization (or shut-in) stage. (c) Details of the porous fault zone near the fluid source. (d,e,f) Distinct stages of rupture propagation in our upper-bound model (see panel (b) indicating the corresponding times as yellow circles). (d) Crack-like rupture phase during the pressurization stage. (e,f) Pulse-like rupture phase during the depressurization stage: first as a ring-like pulse (panel (e)), after as two moon-shaped pulses (panel (f)). See the main text for a detailed description of the stages.
    }
    \label{fig:model-schematics} 
\end{figure}

We consider purely aseismic ruptures nucleated by a localized increase of pore-fluid pressure due to the direct injection of fluids into a porous fault zone of width $w$ (Fig. \ref{fig:model-schematics}c). For simplicity, we begin by examining a fluid injection conducted at a constant volumetric rate $Q_0$ over a finite time $t_s$, followed by a sudden injection stop (Fig. \ref{fig:model-schematics}a). This results in two distinct stages: a continuous injection or pressurization stage, in which pore pressure increases everywhere within the permeable fault zone; and a shut-in or depressurization stage, in which pore pressure decays near the injection point (Fig. \ref{fig:model-schematics}b) while transiently increasing away from it (Fig. \ref{fig:model-schematics}e). Incorporating these two stages into the model allows for examining aseismic ruptures throughout their entire lifetime, from nucleation to arrest. We consider a planar infinite fault obeying a slip-weakening friction law with a static (peak) friction coefficient $f_p$, dynamic (residual) friction coefficient $f_r$, and characteristic slip-weakening distance $\delta_c$. The decay of friction from the peak to the residual value can be either linear or exponential \cite{Saez_Lecampion_2024}. The host rock is considered purely elastic with the same elastic constants as the fault zone (Fig. \ref{fig:model-schematics}c). We assume the host rock to be impermeable at the relevant time scales of the injection. This configuration is motivated by the permeability structure of fault zones in which a highly-permeable damage zone is commonly surrounded by a less permeable host rock \cite{Caine_Evans_1996,Faulkner_Jackson_2010}. Under these assumptions and, 
particularly at large times compared to the characteristic time for diffusion of pore pressure in the direction perpendicular to the fault, $w^2/\alpha$, with $\alpha$ the fault-zone hydraulic diffusivity, the deformation rate in the fault zone is essentially oedometric (uniaxial along the $z$-axis) \cite{Marck_Savitski_2015}. Fluid flow within the fault zone is then governed by an axisymmetric linear diffusion equation for the pore pressure field $p$, $\partial p/\partial t = \alpha \nabla^2p$ \cite{Detournay_Cheng_1993}, where the hydraulic diffusivity $\alpha=k/S\eta$, with $k$ and $\eta$ the fault permeability and fluid dynamic viscosity respectively, and $S$ is the so-called oedometric storage coefficient representing the variation of fluid content caused by a unit pore pressure change under uniaxial strain and constant stress normal to the fault plane \cite{Green_Wang_1990,Detournay_Cheng_1993}. By neglecting any poroelastic coupling within the fault zone upon the activation of slip, deformation in the medium is governed by linear elasticity which by virtue of the slow nature of the slip we are concerned with, is regarded in its quasi-static approximation. Moreover, we assume that fault slip is concentrated in a principal slip zone modeled as a mathematical plane located at $z=0$ (Fig. \ref{fig:model-schematics}c). We presented an investigation of this physical model, which can be regarded as an extension to three dimensions of the two-dimensional plane-strain model of Garagash and Germanovich \cite{Garagash_Germanovich_2012}, in a recent study \cite{Saez_Lecampion_2024}. Here, our main focus is on the case of ruptures that are unconditionally stable according to the terminology and regimes presented in \cite{Saez_Lecampion_2024}. 
This condition requires that the background shear stress $\tau_0$, which is assumed to be uniform, must be lower than the in-situ residual fault strength $f_r\sigma_0^\prime$, where $\sigma_0^\prime=\sigma_0-p_0$, with $\sigma_0$ and $p_0$ the uniform background normal stress and pore pressure respectively. $\tau_0$ and $\sigma_0^\prime$ are thought to be the result of long-term tectonic processes and thus considered to be constant during the times scales associated with the injection operation.

\begin{figure}
    \centering
    \includegraphics[width=16cm]{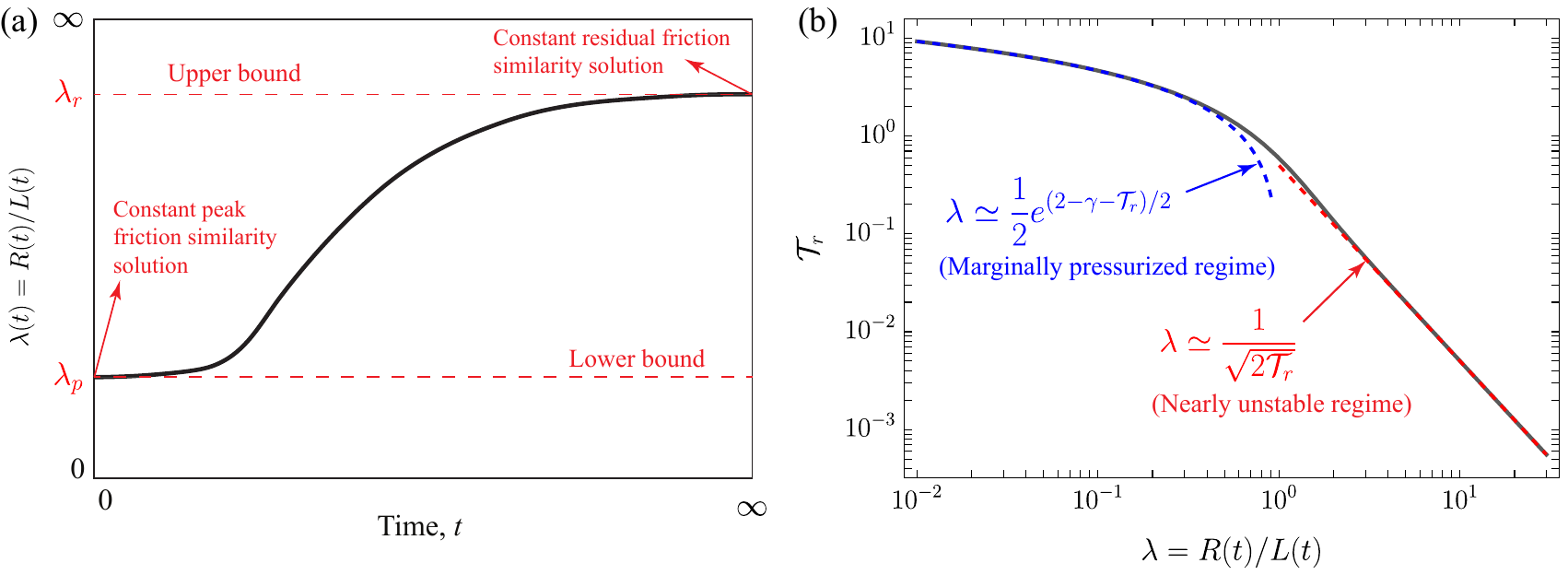}
    \caption{\textbf{Upper bound rationale and amplification factor.} 
    (a) Evolution of the amplification factor $\lambda(t)=R(t)/L(t)$ for unconditionally stable ruptures in the slip-weakening fault model (adapted from figure 9 in \cite{Saez_Lecampion_2024}). $R(t)$ is the rupture radius and $L(t)=\sqrt{4\alpha t}$ is the position of the overpressure front. $\lambda_r$ (time-independent) hereafter denoted simply $\lambda$, is an upper bound for the rupture size at any time during continuous injection. (b) Analytical solution (solid black line) for the amplification factor $\lambda$ in our upper-bound model. $\lambda$ depends uniquely on the residual stress-injection parameter $\mathcal{T}_r$. Blue and red dashed lines correspond to asymptotic limiting behaviors for marginally pressurized ($\lambda\ll1$) and nearly unstable ($\lambda\gg1$) ruptures.}
    \label{fig:upperbound-lambda}
\end{figure}

In our model, unconditionally stable ruptures evolve always between two similarity solutions (Fig. \ref{fig:upperbound-lambda}a and see \cite{Saez_Lecampion_2024} for further details), one at early times where the fault interface operates with a constant friction coefficient equal to the peak value $f_p$, and the other one at late times where the fault interface behaves as if it were governed by a constant friction coefficient equal to the residual value $f_r$. As shown in Fig. \ref{fig:upperbound-lambda}a, the constant residual friction solution, which is the ultimate asymptotic regime of any unconditionally stable rupture, is an upper bound for the rupture size at any given time during the pressurization stage. In this asymptotic regime, rupture growth is dictated by a fracture-mechanics energy balance, where the interplay between driving and resisting forces leads to a scenario in which the fracture energy can be effectively neglected \cite{Saez_Lecampion_2024}. For this reason, this upper-bound limiting regime is also referred to as the zero-fracture-energy solution \cite{Saez_Lecampion_2024}. During the shut-in stage, a similar upper-bound rationale can be applied to the case of unconditionally stable ruptures. Assuming the fault slides with a constant residual friction value across the slipping region (equivalent to neglecting the fracture energy in the rupture-tip energy balance), this limiting solution would consistently yield a maximum for the rupture size, as the effect of the fracture energy is always to slow down rupture advancement. The limiting scenario of constant residual friction serves, therefore, as an effective upper-bound model for unconditionally stable ruptures from nucleation to arrest. In the following sections, we explore the consequences of such a limiting condition to provide an upper bound for the
evolution of the rupture size and moment release during and after fluid injection, as well as
a theoretical estimate for the final, maximum size and magnitude of injection-induced slow slip events. While the canonical example of injection at a constant flow rate is used to examine injection-induced aseismic slip in a relatively comprehensive manner which includes different stages and regimes, we emphasize in advance that our estimates for the maximum size and magnitude will account for fluid injections conducted with an arbitrary volumetric rate history (Fig. \ref{fig:model-schematics}a).

\subsection{Dynamics of unconditionally stable ruptures and maximum rupture size}\label{sec:rupture-size}

During and after fluid injection, our upper-bound model is governed by a single dimensionless number (see Methods), the so-called residual stress-injection parameter:
\begin{equation}\label{eq:residual-stress-injection-parameter}
    \mathcal{T}_r= \frac{\Delta\tau_{r-0}}{f_r\Delta p_*},\quad\text{with}\quad \Delta\tau_{r-0}=f_{r}\sigma_{0}^{\prime}-\tau_{0} \quad \text{and}\quad \Delta p_*=Q_0\eta/4\pi kw.
\end{equation}
This dimensionless number systematically emerges in physics-based models of injection-induced fault slip \cite{Bhattacharya_Viesca_2019,Saez_Lecampion_2022,Saez_Lecampion_2023,Ciardo_Lecampion_2023,Saez_Lecampion_2024}. It quantifies the competition between the two opposite \textit{forces} that determine the dynamics of unconditionally-stable ruptures in our upper-bound model. One is a driving force associated with the sole effect of pore pressure increase due to fluid injection which continuously reduces fault shear strength, thereby releasing elastic strain energy that becomes available for rupture growth. Its stress scale is $f_r\Delta p_*$, where $\Delta p_*$ is the injection intensity. Injections with faster pressurization are associated with increasing values of $\Delta p_*$ which can occur, for example, due to a higher injection flow rate ($Q_0$) or a lower hydraulic transmissivity ($kw$). The other force is of a resisting kind, which in the absence of a local energy dissipation mechanism such as the fracture energy, corresponds to a non-local \textit{consumption} of elastic strain energy associated with the background stress change, $\Delta\tau_{r-0}$. The latter is defined as the difference between the in-situ residual fault strength ($f_r\sigma_0^\prime$) and the initial shear stress ($\tau_0$). The former quantity can be also interpreted as the final shear stress that would act on the slipped fault patch after the termination of the injection operation and the subsequent dissipation of overpressure due to the injection. Hence, $\Delta\tau_{r-0}$ quantifies a change of shear stress between a final and initial state. The background stress change is strictly positive. This is an essential feature of unconditionally stable ruptures. Specifically, the ultimate stability condition: $\tau_0<f_r\sigma_0^\prime$, ensures the fault stability and the development of quasi-static slip unconditionally in this regime \cite{Garagash_Germanovich_2012,Saez_Lecampion_2024}. Intuitively one expects that as the intensity of the injection ($\Delta p_*$) increases, the rupture would propagate faster. Conversely, when the background stress change ($\Delta\tau_{r-0}$) is higher, it presents greater resistance to rupture growth, consequently slowing down the slip propagation. Hence, decreasing $\mathcal{T}_r$ values will always result in faster aseismic ruptures. This behavior can be clearly observed in Fig. \ref{fig:upperbound-lambda}b, where the solution (see Methods) during the pressurization stage for a circular rupture of radius $R(t)$ is shown. Here, $R(t)=\lambda L(t)$, where $\lambda$ is the so-called amplification factor \cite{Bhattacharya_Viesca_2019}, and $L(t)=\sqrt{4\alpha t}$ is the classical diffusion length scale, also considered as the nominal position of the overpressure front (Fig. \ref{fig:model-schematics}d). $\lambda$ therefore relates the position of the overpressure and slip fronts. This analytical circular-rupture solution is strictly valid only when $\nu=0$ \cite{Saez_Lecampion_2022}. Throughout this work, we generally adopt the circular rupture approximation to derive purely analytical insights. We, nevertheless, quantify the effect of rupture non-circularity numerically via a boundary-element-based numerical solver (see Methods). 

The analytical solution in Fig.~\ref{fig:upperbound-lambda}b provides important insights into the response of our upper-bound model. During the pressurization stage, the fault response is characterized by two distinct regimes. When $\mathcal{T}_r\sim10$, aseismic ruptures are confined well within the overpressurized region ($\lambda\ll1$), a regime known as marginally pressurized because it relates to a scenario in which the fluid injection provides just the minimum amount of overpressure that is necessary to activate fault slip \cite{Garagash_Germanovich_2012,Saez_Lecampion_2024}. Conversely, when $\mathcal{T}_r\ll1$, aseismic ruptures break regions much further away than the pressurized fault zone ($\lambda\gg1$). This is the so-called nearly unstable regime \cite{Saez_Lecampion_2024} as when $\Delta \tau_{r-0}\to 0$ the rupture approaches the condition under which it becomes ultimately unstable. From a practical standpoint and in an upper-bound sense, this is indeed the most relevant regime as it produces the largest ruptures for a given injection. While operators in geo-energy applications typically maintain good control over the parameters of the fluid injection, in-situ conditions such as the stress state acting upon fractures and faults within a reservoir are subject to significant uncertainties. Given that in-situ conditions largely control the response of aseismic slip in our model, it seems reasonable to assume under rather generic, generally uncertain conditions in the rock mass surrounding a given operation, that the nearly unstable regime provides an upper limit for the size and magnitude of aseismic slip events. Consequently, our emphasis in this work will predominantly be on exploring this regime. In fact, when $\lambda\gg1$, one can derive a relation linking the evolution of the rupture radius to the accumulated injected fluid volume ($V(t)$) and in-situ conditions as follows (see Methods):
\begin{equation}\label{eq:R-volume-Q-const}
    R(t)=A_\text{situ}\sqrt{V(t)},\quad \text{with}\quad A_\text{situ}=\left(\frac{f_r}{2\pi wS \Delta\tau_{r-0}}\right)^{1/2}.
\end{equation}
This equation is valid not only for injection at a constant flow rate but also for any arbitrary fluid injection as long as the rupture propagates in crack-like mode during the pressurization stage (see Methods), so that the fracture-mechanics energy balance, equation \eqref{eq:energy-balance}, remains valid. A crack-like propagation mode will certainly hold at least in one relevant scenario, wherein overpressure due to fluid injection increases monotonically everywhere within the sliding region. Hereafter, to put the term arbitrary in a more specific but still sufficiently general scope, we refer to injection with monotonically increasing fluid pressure as \textit{arbitrary}. However, we emphasize that in our model, a monotonically increasing fluid pressure is only a sufficient (not a necessary and sufficient) condition for crack-like propagation. Moreover, equation \eqref{eq:R-volume-Q-const} implies that, during the pressurization stage, the cumulative injected fluid volume $V(t)$ is the only operational parameter of the injection that matters to estimate an upper bound for the rupture size at a given time $t$. Furthermore, the prefactor is exclusively related to in-situ conditions ($A_\text{situ}$).

\begin{figure}
    \centering
    \includegraphics[width=16cm]{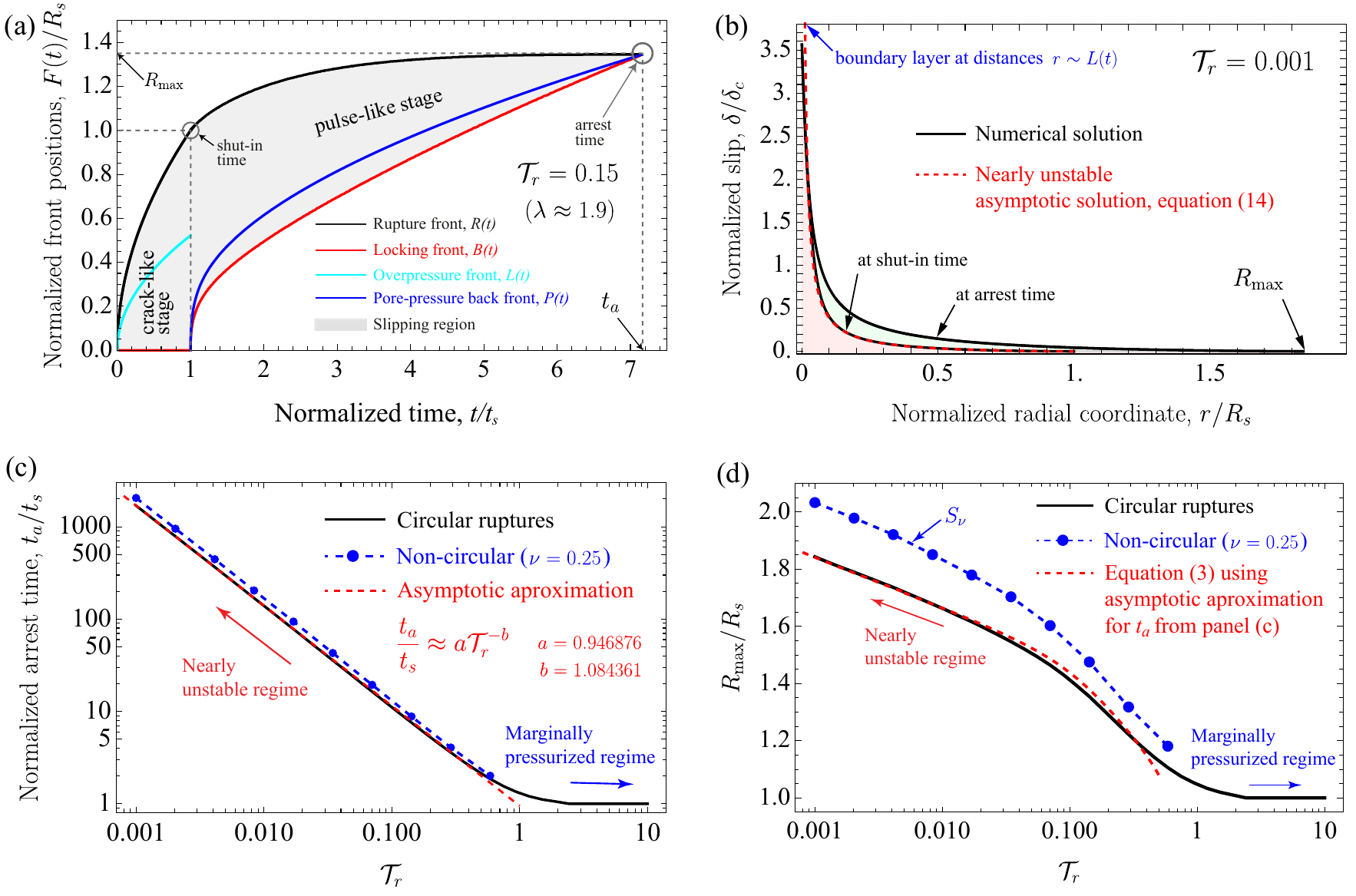}
    \caption{\textbf{Dynamics of unconditionally stable ruptures, arrest time, and maximum rupture size.} 
    (a) Evolution of fluid- and slip-related fronts, during the entire lifetime of an injection-induced aseismic slip event, for a case with $\mathcal{T}_r=0.15$. Front positions $F(t)=R(t),B(t),L(t),P(t)$ are normalized by the rupture radius at the shut-in time, $R_s$ (see the main text for a description of the different fronts). (b) Slip distribution for a very nearly unstable rupture ($\mathcal{T}_r\ll 1$) at the shut-in and arrest times. 
    $\delta_c(t)=f_r\Delta p_* L(t)/\mu$
    is the characteristic slip scale in this regime (see Methods). Slip is further accumulated during the shut-in stage due to the slip pulse that travels along the fault upon the shut-in of the injection. (c) Upper bound for the arrest time and (d) maximum rupture run-out distance as a function of the dimensionless parameter $\mathcal{T}_r$.}
    \label{fig:maximum-size}
\end{figure}

Upon the stop of the fluid injection ($t>t_s$), our upper-bound model produces ruptures that transition from crack-like to pulse-like propagation mode (Fig. \ref{fig:model-schematics}d-f). Indeed, since we have reduced the upper-bound problem to a fault responding with a constant friction coefficient equal to the residual value $f_r$, we inherit essentially all the results obtained recently by Sáez and Lecampion \cite{Saez_Lecampion_2023} who extensively investigated the propagation and arrest of post-injection aseismic slip on a fault with constant friction. In particular, the overpressure drops quickly near the fluid source upon the stop of the injection while it keeps increasing transiently away from it (Fig.~\ref{fig:model-schematics}d-e). This latter increase of pore pressure is what further drives the propagation of aseismic ruptures after shut-in. As shown in Fig.~\ref{fig:model-schematics}e, slip propagates first as a ring-shaped pulse with a locking front that propagates always faster than the rupture front (Fig.~\ref{fig:maximum-size}a). The locking front is driven by the continuous depressurization of pore fluids which re-strengthens the fault. After, and for the more general case of non-circular ruptures, the pulse splits into two `moon-shaped' pulses (Fig. \ref{fig:model-schematics}f). This ultimate stage is due to the locking front catching up with the rupture front first in the less elongated side of the slipping region. For the idealized case of circular ruptures, the moon-shaped pulses are absent due to the axisymmetry property of both the fluid flow and shear rupture problems. Fig. \ref{fig:maximum-size}a displays the evolution of the locking front $B(t)$ and rupture front $R(t)$ for a circular rupture for an exemplifying case with $\mathcal{T}_r=0.15$. Slip arrests when the locking front catches the rupture front at the time $t_a$ (arrest time or duration of the slow slip event), resulting in the maximum rupture run-out distance $R_\text{max}$. Although perhaps more insightfully, the rupture front stops when it is caught by the so-called pore-pressure back front $P(t)$ \cite{Saez_Lecampion_2023} introduced by Parotidis \textit{et al.} \cite{Parotidis_Shapiro_2004}. This latter means that there is no further increase of pore pressure within the rupture pulse that is available to sustain the propagation of slip. Moreover, this arrest condition leads to the following analytical relation between the maximum rupture radius $R_\text{max}$ and the arrest time $t_a$:
\begin{equation}\label{eq:ta-Ra}
    R_\text{max}=\left[4\alpha t_a\left(\frac{t_a}{t_s}-1\right)\ln\left(\frac{t_a}{t_a-t_s}\right)\right]^{1/2}.
\end{equation}
In the more practically relevant, nearly unstable regime ($\mathcal{T}_r\ll 1$), the normalized arrest time (Fig. \ref{fig:maximum-size}c) can be estimated via the following numerically-derived asymptotic approximation, $t_a/t_s\approx a\mathcal{T}_r^{-b}$, with $a=0.946876$ and $b=1.084361$. Moreover, Fig. \ref{fig:maximum-size}c shows that when ruptures are marginally pressurized ($\mathcal{T}_r\sim10$), the slip pulses arrest almost immediately after the injection stops. Conversely, when ruptures are nearly unstable ($\mathcal{T}_r\ll1$), the upper bound for the arrest time ($t_a$) is predicted to be several orders of magnitude the injection duration ($t_s$). Rupture non-circularity has the effect of slightly increasing both the arrest time and maximum rupture run-out distance $R_\text{max}$ (Figs. \ref{fig:maximum-size}c-d). Furthermore, the contribution of the shut-in stage to $R_\text{max}$ is approximately a factor of two at most when ruptures are very nearly unstable ($\mathcal{T}_r\sim0.001$, Fig. \ref{fig:maximum-size}d). Hence, the order of magnitude of $R_\text{max}$ comes directly from evaluating $R(t)$ in the analytical solution displayed in Fig \ref{fig:upperbound-lambda}b at the shut-in time, which in the regime $\lambda\gg1$ takes a more insightful expression given by equation \eqref{eq:R-volume-Q-const}, which is valid for arbitrary fluid injections. Using this latter expression, we can calculate the maximum run-out distance when $\lambda\gg1$ as:
\begin{equation}\label{eq:Rmax-nearly-unstable}
    R_\text{max}=S_\nu A_\text{situ}\sqrt{V_\text{tot}},
\end{equation}
where $V_\text{tot}=V(t_s)$ is the total volume of fluid injected during a given operation, and the coefficient $S_\nu$ accounts for the further growth of the rupture during the shut-in stage and the effect of rupture non-circularity. $S_\nu$ is a function of $\mathcal{T}_r$ and $\nu$ and can be simply approximated by the blue dashed line in Fig. \ref{fig:maximum-size}d for $\nu=0.25$.


\subsection{Maximum moment release and magnitude}\label{sec:moment-release-magnitude}

\begin{figure}
    \centering
    \includegraphics[width=17.0cm]{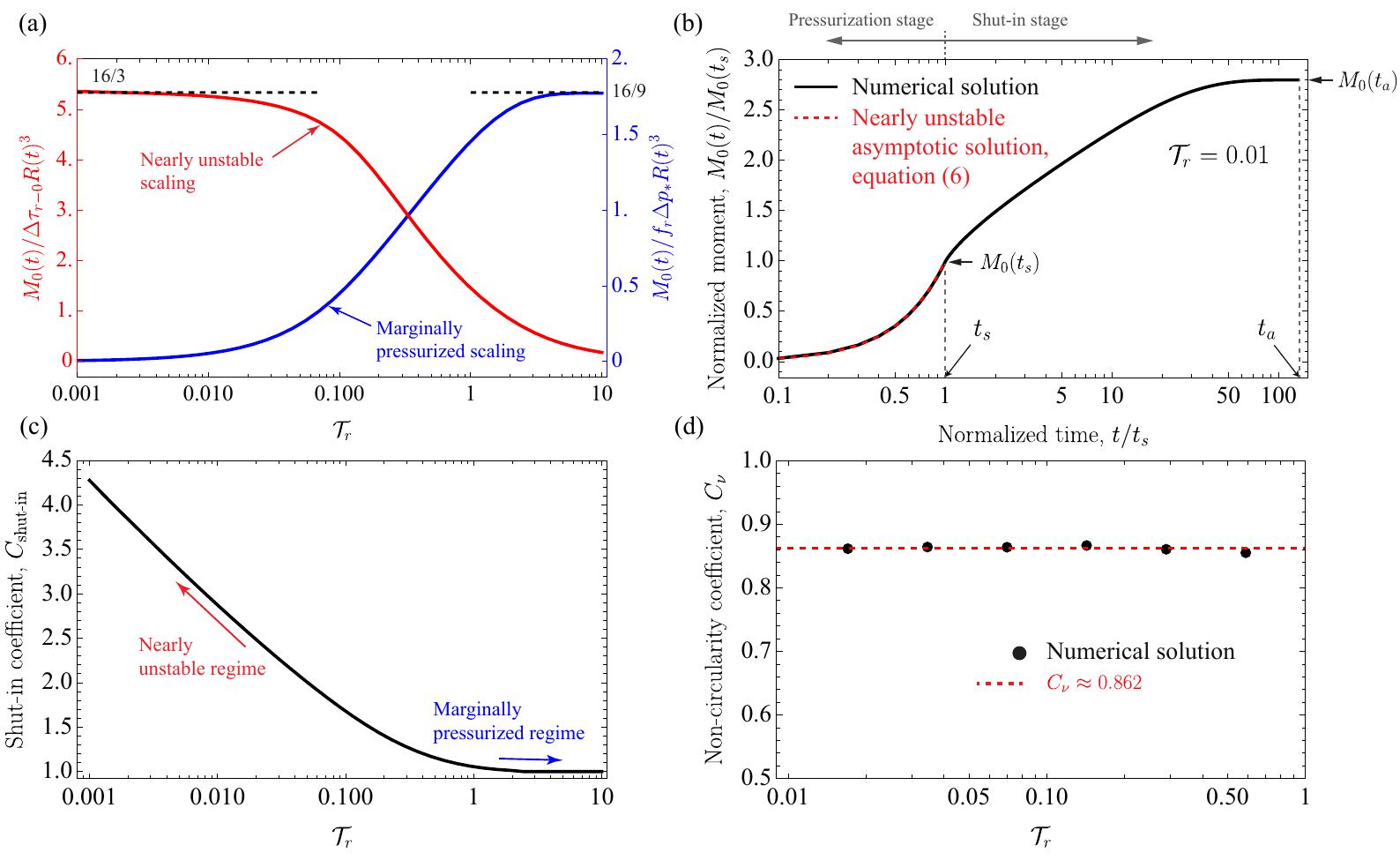}
    \caption{\textbf{Upper-bound model for the moment release.} (a) Normalized moment release during the pressurization stage as a function of the residual stress-injection parameter $\mathcal{T}_r$ using the nearly unstable (red, left axis) and marginally pressurized (blue, right axis) scalings. Black dashed lines correspond to asymptotic analytical solutions provided in the main text. Solid lines correspond to numerical solutions. (b) Evolution of the moment release 
    for a nearly unstable circular rupture ($\mathcal{T}_r\ll 1$) during and after fluid injection. The moment release increases up to $\approx$ 3 times the moment release at the shut-in time in this particular case. (c) Shut-in coefficient $C_\text{shut-in}=M_0(t_a)/M_0(t_s)$ for a circular rupture as a function of the dimensionless parameter $\mathcal{T}_r$. Nearly unstable ruptures can experience a moment release increment of up to $\approx$ 4 times during the shut-in stage. Conversely, marginally pressurized ruptures ($\mathcal{T}_r\sim 10$) are characterized by no increment at all. (d) Non-circularity coefficient $C_\nu$ as a function of $\mathcal{T}_r$ for $\nu=0.25$. The numerical simulations suggest that over a wide range of practically relevant cases ($0.01\leq\mathcal{T}_r\leq 1$), the moment release of a non-circular rupture is about 13.8 percent smaller than the moment release of a circular rupture ($\nu=0$) at same $\mathcal{T}_r$.}
    \label{fig:maximum-moment}
\end{figure}

To calculate the moment release, we derive analytical upper bounds for the spatiotemporal evolution of fault slip during the pressurization stage, for both nearly unstable ($\lambda\gg1$) and marginally pressurized ($\lambda\ll1$) ruptures (see Methods). Notably, the slip distribution of nearly unstable ruptures is highly concentrated around the injection point due to a boundary layer associated with the fluid-injection force at distances $r\sim L(t)$ (Fig. \ref{fig:maximum-size}b). Upon integrating the analytical slip distributions over the rupture surface, the temporal evolution of moment release is:
\begin{equation}\label{eq:moment-pressurization-rupture-radius}
    M_0\simeq
    \begin{cases}
        \frac{16}{3}\Delta \tau_{r-0}R^3 & \text{for nearly unstable ruptures, }\lambda\gg1,\\
        \frac{16}{9}f_r\Delta p_* R^3 & \text{for marginally pressurized ruptures, } \lambda\ll1,
    \end{cases}
\end{equation}
with the temporal dependence of $M_0$ embedded implicitly in $R(t)=\lambda L(t)$ which is known analytically (Fig. \ref{fig:upperbound-lambda}b). As expected, the previous asymptotic solutions for $M_0$ match very closely the full numerical solution (Fig. \ref{fig:maximum-moment}a). The numerical solution helps us to describe the precise transition between the two end members. We emphasize that the structure of the scaling for $M_0$ is evidently the one expected for a circular crack ($M_0\propto R^3$). Yet the pre-factors and relevant stress scales are specific to the characteristic loading of each regime. For instance, in the nearly unstable regime ($\lambda\gg 1$), the proper stress scale is the background stress change ($\Delta \tau_{r-0}$), as opposed to the injection intensity ($f_r\Delta p_*$) which is the adequate stress scale when $\lambda\ll1$. This is because, in the nearly unstable regime, most of the slipping region experiences a uniform stress variation $\Delta \tau_{r-0}$ except for a very small region of size $\sim L(t)$ near the fluid source which undergoes an additional non-uniform stress change due to the fluid injection. The effect of the fluid-injection force is indeed in the pre-factor 16/3, which is about two times bigger than the one of a circular crack with purely uniform stress drop (16/7 when $\nu=0.25$ \cite{Kanamori_Anderson_1975}, and 8/3 when $\nu=0$ \cite{Segedin_1951}). Moreover, in this regime, we obtain the following expression for the moment release which is valid for arbitrary fluid injections (see Methods):
\begin{equation}\label{eq:moment-volume-constant-rate}
    M_0(t)= I_\text{situ} \cdot V(t)^{3/2}, \quad \text{with} \quad I_\text{situ}=\frac{16}{3(2\pi)^{3/2}}\frac{1}{\sqrt{\Delta\tau_{r-0}}}\left( \frac{f_r}{wS} \right)^{3/2}.
\end{equation}
Equation \eqref{eq:moment-volume-constant-rate} has the same property as equation \eqref{eq:R-volume-Q-const}, that is, the only operational parameter of the injection controlling the upper bound for the moment release during the pressurization stage is the cumulative injected fluid volume $V(t)$. Furthermore, the prefactor corresponds as well to in-situ conditions ($I_\text{situ}$), thus effectively separating contributions to the moment release that are controllable during an operation ($V$) and those that are not ($I_\text{situ}$). Such kind of relation for the moment release has been previously reported in the literature for the case of regular, fast earthquakes \cite{Shapiro_Kruger_2011,McGarr_2014,vanderElst_Page_2016,Galis_Ampuero_2017,Li_Elsworth_2021}.

Equation \eqref{eq:moment-volume-constant-rate} shows that a decrease in background stress change ($\Delta \tau_{r-0}$) leads to an increase in moment release. The reason behind such behavior is simple, lower background stress variations result in less opposition for the rupture to grow and thus in a higher moment release. For the same reason, higher values of the residual friction coefficient ($f_r$) also augment $M_0$. On the other hand, decreasing the product between the fault-zone width and oedometric storage coefficient ($wS$), hereafter denominated as fault-zone storativity, also leads to a larger moment release. The explanation, in this case, is that $wS$ controls the pressurization intensity due to fluid injection that is experienced on average over the fault pressurized region. A lower storativity in the fault zone naturally implies a higher fluid pressure to accommodate a fixed amount of injected volume (see equation \eqref{eq:volume-integral}). A higher fluid overpressure decreases fault shear strength therefore increasing the mechanical energy available for rupture growth and the corresponding moment release. It is important to note that in the marginally pressurized regime ($\lambda\ll1$), $M_0$ does not follow an expression as in \eqref{eq:moment-volume-constant-rate}. Indeed, by substituting the expressions $\Delta p_*=Q_0\eta/4\pi kw$, $R(t)=\lambda\sqrt{4\alpha t}$, and $V(t)=Q_0 t$ into equation \eqref{eq:moment-pressurization-rupture-radius}, one can readily show that the moment release for an injection at a constant flow rate is given by $M_0(t)= B \cdot V(t)^{3/2}$, with $B=(32/9\pi)(f_r\eta/kw)(\lambda^3\alpha^{3/2}/Q_0^{1/2})$. This implies that in this regime, the moment release depends on both the current injected volume $V(t)$ (or injection time $t$) and the injection rate $Q_0$ (which is also implicitly in $\lambda$). More importantly, the in-situ and operational factors cannot be separated as in \eqref{eq:moment-volume-constant-rate}. This separation is a unique characteristic of nearly unstable ruptures, associated with the fact that when $\lambda\gg1$, the effect of the fluid source on rupture propagation is entirely described by its equivalent force at distances $r\gg L(t)$. The magnitude of this equivalent force is determined only by the injected fluid volume, irrespective of any other detail of the fluid injection (see Methods, equation \eqref{eq:point-force-definition}).

During the shut-in stage ($t>t_s$), the propagation and ultimate arrest of the aseismic slip pulses result in a further accumulation of fault slip (Fig. \ref{fig:maximum-size}b). The depressurization stage thus increases the final, maximum moment release of the events. Fig. \ref{fig:maximum-moment}b displays the evolution of this increase for an exemplifying case with $\mathcal{T}_r=0.01$. We observe that the moment release keeps growing after shut-in very slowly (over a timescale that is about 100 times the injection duration) up to reaching (at arrest) nearly three times the moment release at the time the injection stops ($M_0(t_s)$). We quantify this effect in the most general form by defining the shut-in coefficient $C_\text{shut-in}$, equal to the ratio between the maximum moment release at the time in which a circular rupture arrest, $M_0(t_a)$, and the moment release at the shut-in time, $M_0(t_s)$. By dimensional analysis, the shut-in coefficient depends only on the residual stress-injection parameter $\mathcal{T}_r$, whose relation is calculated numerically and displayed in Fig. \ref{fig:maximum-moment}c. We observe that $M_0(t_a)$ is at most around 4 times the moment release at the time the injection stops in the more nearly unstable cases (smallest values of $\mathcal{T}_r$). Conversely, there is virtually no further accumulation of moment release for marginally pressurized ruptures. We quantify the effect of rupture non-circularity in a similar way by introducing the coefficient $C_\nu$ equal to the ratio between the moment release at the time of arrest for non-circular ruptures ($\nu\neq0$), and the same quantity for the circular case ($\nu=0$). Again, by dimensional considerations, $C_\nu$ depends only on $\mathcal{T}_r$ for a given $\nu$. This is shown in Fig.~\ref{fig:maximum-moment}d for the particular case of a Poisson's solid ($\nu=0.25$, a common approximation for rocks). We observe that the effect of the Poisson's ratio is to reduce in about 13.8 percent the moment release of a non-circular rupture with respect to the one of a circular rupture, for the same $\mathcal{T}_r$. We find this to be valid over a wide range of practically relevant cases ($0.01\leq\mathcal{T}_r\leq 1$). With all the previous definitions and calculations, we can finally estimate the maximum moment release as $M_0^\text{max}= C_\nu \cdot C_\text{shut-in} \cdot M_0(t_s)$. Notably, in the nearly unstable regime ($\lambda \gg 1$), equation \eqref{eq:moment-volume-constant-rate} can be evaluated at the shut-in time, which allows us to arrive at the following expression valid for arbitrary fluid injections:
\begin{equation}\label{eq:final-moment-volume-constant-rate}
    M_0^\text{max}= C_\nu \cdot C_\text{shut-in} \cdot I_\text{situ} \cdot V_\text{tot}^{3/2},
\end{equation}
where $V_\text{tot}=V(t_s)$ is the total volume of fluid injected during a given operation. Equation \eqref{eq:final-moment-volume-constant-rate} has a multiplicative form, thus effectively factorizing contributions from the injected fluid volume, in-situ conditions, shut-in stage, and rupture non-circularity to the maximum moment release. Note that both $C_\text{shut-in}$ and $C_\nu$ depend on $\mathcal{T}_r$ and thus also on in-situ conditions and parameters of the injection (equation \eqref{eq:residual-stress-injection-parameter}). However, the in-situ conditions and injection protocol are for the most part contained in $I_\text{situ}$ and $V_\text{tot}$, respectively, which can vary over several orders of magnitude. On the contrary, the dimensionless coefficients $C_\text{shut-in}$ and $C_\nu$ remain always of order one. Moreover, we shall keep in mind that these latter two coefficients are, strictly speaking, defined only for injection at a constant flow rate. Nevertheless, one could crudely approximate any other kind of injection protocol as $Q_\text{eq}=(1/t_s)\int_0^{t_s} Q(t)\text{d}t$ for the purpose of estimating these two coefficients. The previous approximation guarantees that the same amount of fluid volume is injected over the same injection period $t_s$ by both the equivalent constant-rate source $Q_\text{eq}$ and the time-varying arbitrary source $Q(t)$. Finally, for marginally pressurized ruptures ($\lambda\ll1$), a similar expression for the maximum moment release can be derived as $M_0^\text{max}= C_\nu \cdot B \cdot V_\text{tot}^{3/2}$ (since $C_\text{shut-in}\approx1$, Fig. \ref{fig:maximum-moment}c). As already discussed, the in-situ and operational factors cannot be separated in this regime.

To calculate the maximum magnitude, we follow the definition by Hanks and Kanamori \cite{Hanks_Kanamori_1979}: $M_w^\text{max}=2/3\cdot[\text{log}_{10}\left(M_0^\text{max}\right)-9.1]$ (here, in SI units). In the regime that provides the largest rupture size and moment release for a given injection ($\lambda\gg1$), equation \eqref{eq:final-moment-volume-constant-rate} leads to the following estimate for the maximum magnitude:
\begin{equation}\label{eq:final-maximum-magnitude}
    M_w^\text{max}=\text{log}_{10}\left(V_\text{tot}\right) + \frac{2}{3}\left[\text{log}_{10}\left(I_\text{situ}\right) + \text{log}_{10}\left(C_\text{shut-in}\right) + \text{log}_{10}\left(C_\nu\right) - 9.1 \right].
\end{equation}

Due to the multiplicative form of equation \eqref{eq:final-moment-volume-constant-rate}, equation \eqref{eq:final-maximum-magnitude} takes an additive form that separates contributions from different factors to the maximum magnitude of injection-induced slow slip events. 
Among these factors, rupture non-circularity decreases the magnitude only by 0.06. The contribution from the shut-in stage is, on the other hand, slightly larger. Since $C_\text{shut-in}\approx 4$ at most when ruptures are very nearly unstable ($\mathcal{T}_r\sim0.001$), the shut-in stage may contribute to an increase in the moment magnitude of $0.4$ at the maximum. The larger contributions to $M_w^\text{max}$ are by far the ones associated with in-situ conditions and the total injected fluid volume. For example, a tenfold increase in $V_\text{tot}$ gives a magnitude increase of 1.0, while a tenfold increase in $I_\text{situ}$ results in a magnitude growth of approximately $0.67$. The relative contributions from the sub-factors composing $I_\text{situ}$ can be further understood by substituting equation \eqref{eq:moment-volume-constant-rate} into \eqref{eq:final-maximum-magnitude}, and then isolating the in-situ term as follows:
\begin{equation}\label{eq:in-situ-contributions}
    (2/3)\text{log}_{10}\left(I_\text{situ}\right)=\text{log}_{10}\left(f_r\right)-(1/3)\text{log}_{10}\left(\Delta \tau_{r-0}\right)-\text{log}_{10}\left(wS\right)-0.3135.
\end{equation}
The more significant variations in $M_w^\text{max}$ come clearly from the fault-zone storativity ($wS$) and background stress change ($\Delta \tau_{r-0}$), which could vary over several orders of magnitude. For instance, a variation of three orders of magnitude in $\Delta \tau_{r-0}$ yields a change of magnitude of 1.0, while the same variation in fault-zone storativity results in a magnitude change of 3.0, highlighting the potentially strong effect of $wS$ in $M_w^\text{max}$.

\begin{figure}
    \centering
    \includegraphics[width=16cm]{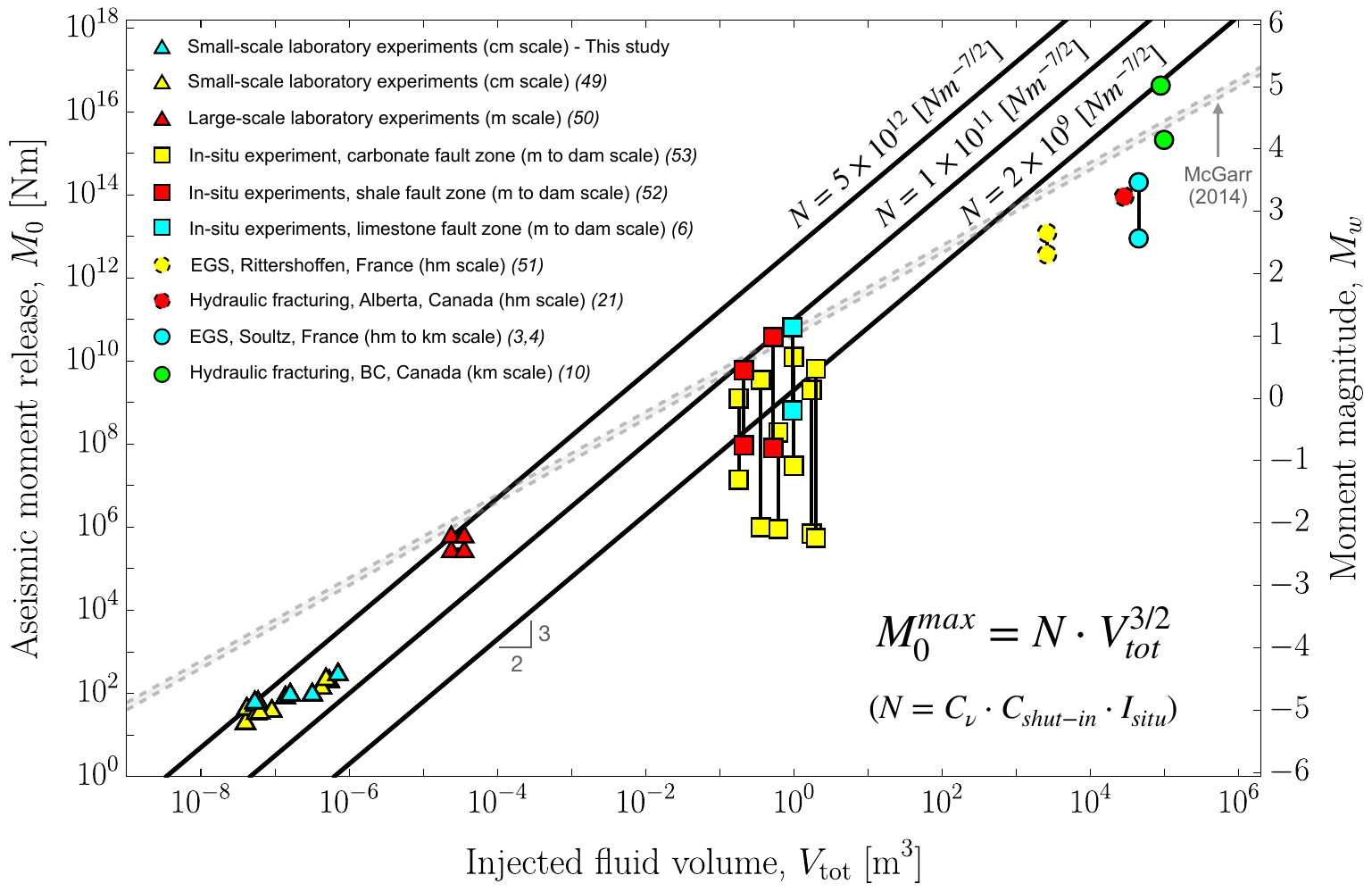}
    \caption{\textbf{Comparison of our scaling relation for the maximum magnitude $M_w^\text{max}$ with estimates of moment magnitude from injection-induced slow slip events, as a function of the total injected fluid volume.} We consider three values of the factor $N$ (solid black lines) which collectively form an upper bound for the data across different volume and moment release scales. For some events in the dataset, the moment release is estimated within a range that is represented by a vertical line connecting their maximum and minimum values (see Supplementary Materials for further details). Gray dashed lines represent McGarr's relation \cite{McGarr_2014} for shear moduli of 20 and 30 GPa.}
    \label{fig:moment-observations}
\end{figure}

\subsection{Fault-zone storativity and injected fluid volume: two key parameters}\label{sec:storativity-importance}

\begin{figure}
    \centering
    \includegraphics[width=16cm]{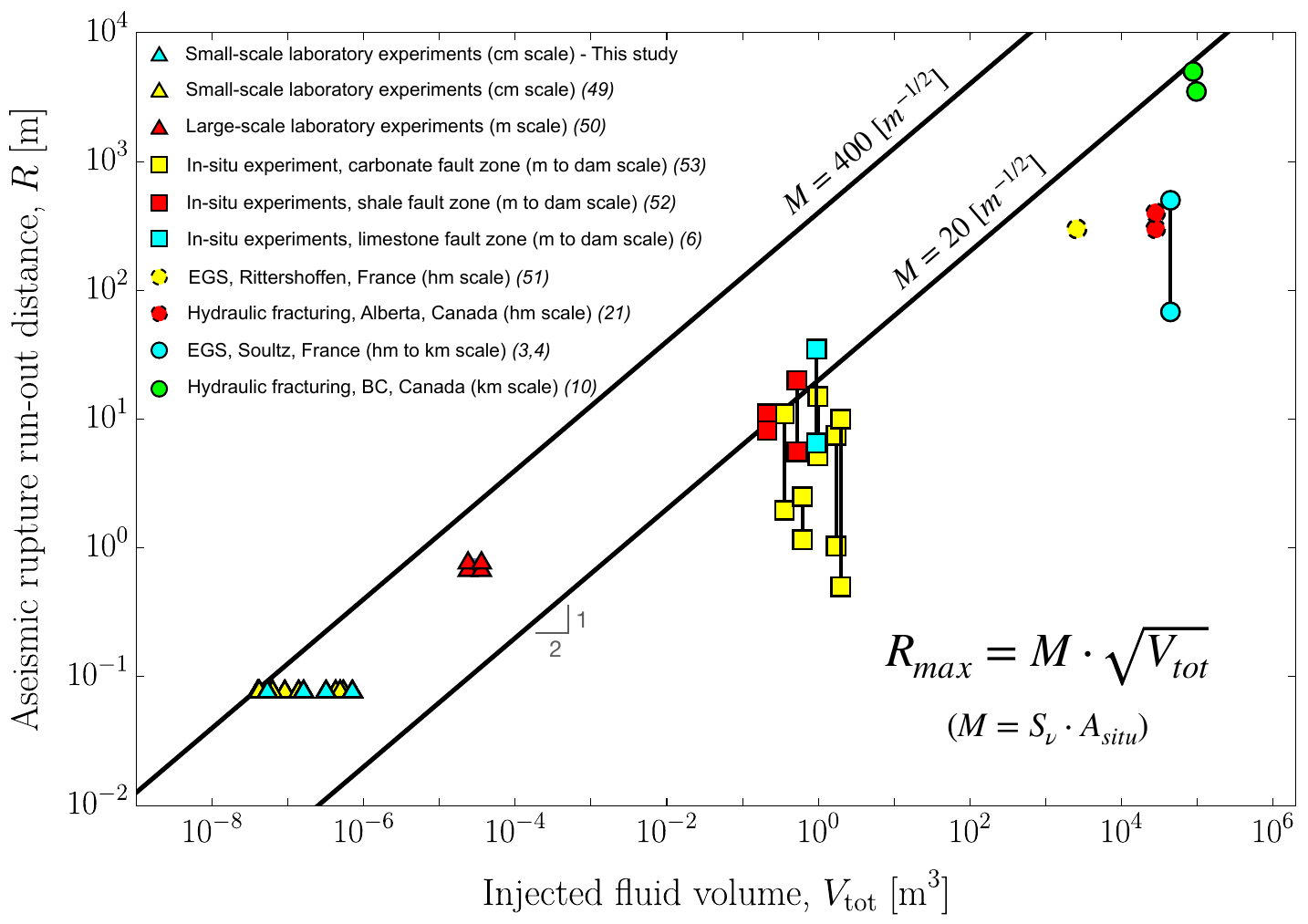}
    \caption{\textbf{Comparison of our scaling relation for the maximum rupture run-out distance $R_\text{max}$ with estimates of rupture extent for the same injection-induced slow slip events as in Fig. \ref{fig:moment-observations}, as a function of the total injected fluid volume.} We consider two values of the factor $M$ (solid black lines) which together form an upper bound for the data across different volume and rupture run-out distance scales. Likewise in Fig. \ref{fig:moment-observations}, for some events the rupture extent is estimated within a range that is represented by a vertical line connecting their maximum and minimum values (see Supplementary Materials for further details)
    }
    \label{fig:size-observations}
\end{figure}

To test our scaling relations, we compiled and produced a new dataset (Supplementary Materials) with estimates of aseismic moment release, rupture size, and injected fluid volumes from events that vary in size from laboratory experiments (centimetric to metric scale ruptures) \cite{Passelegue_Almakari_2020,Cebry_Ke_2022} to industrial applications (hectometric to kilometric scale ruptures) \cite{Cornet_Helm_1997,Bourouis_Bernard_2007,Lengline_Boubacar_2017,Eyre_Eaton_2019,Eyre_Samsonov_2022}, including in-situ experiments in shallow natural faults at intermediate scales (metric to decametric ruptures) \cite{Guglielmi_Cappa_2015,DeBarros_Daniel_2016,Duboeuf_DeBarros_2017}. The comparison between this dataset and our expressions for the maximum moment release \eqref{eq:final-moment-volume-constant-rate} (or magnitude \eqref{eq:final-maximum-magnitude}) and maximum rupture size \eqref{eq:ta-Ra}, are displayed in Figs.~\ref{fig:moment-observations} and \ref{fig:size-observations} respectively. We focus first on the maximum moment release (Fig.~\ref{fig:moment-observations}). To facilitate the comparison against the dataset, we introduce in Fig.~\ref{fig:moment-observations} the factor $N=C_\nu \cdot C_\text{shut-in} \cdot I_\text{situ}$ which encapsulates all effects other than the injected fluid volume, so that equation \eqref{eq:final-moment-volume-constant-rate} can be simply written as $M_0^\text{max}=N\cdot V_\text{tot}^{3/2}$. Three different values for $N$ are considered in Fig. \ref{fig:moment-observations} which collectively form an upper bound for the data across the different volume and moment release scales characterizing the dataset. Considering that $C_\nu\approx 0.862$ and that plausible values for the coefficient $C_\text{shut-in}$ range from 1 to 4, the order of magnitude and units of $N$ are the ones determined by the in-situ factor ($I_\text{situ}$)
. This latter, in turn, depends on three parameters: the residual friction coefficient $f_r$ (with a plausible range of 0.4 to 0.8), the background stress change ($\Delta \tau_{r-0}$), and the fault-zone storativity ($wS$). The background stress change can be at most equal to the amount of shear stress that is necessary to activate fault slip before the injection starts, $\Delta \tau_{p-0}=f_p\sigma_0^\prime-\tau_0$, in the limiting case in which the weakening of friction is small ($f_r\approx f_p$). Its minimum value could be, on the other hand, as small (but positive) as possible when the residual fault strength drops close to the initial shear stress ($f_r\sigma_0^\prime\approx\tau_0$). This is, as already discussed, the case that would promote larger ruptures and moment release. $\Delta \tau_{r-0}$ could therefore reasonably fluctuate between some MPa and a few kPa. The fault-zone storativity ($wS$) may similarly vary over several or potentially many orders of magnitude \cite{Kuang_Jiao_2020,Doan_Brodsky_2006,Xue_Brodsky_2013,Rutqvist_Noorishad_1998}. Estimating this parameter is quite challenging; however, as anticipated by equation \eqref{eq:in-situ-contributions}, $wS$ could have a strong effect on the maximum magnitude. Hence, we conduct a more intricate analysis of representative values for $wS$ within our compilation of events.

To do so, we examine the end members of our data points, namely, small-scale laboratory experiments and industrial-scale fluid injections. Let us first note that in our model, $wS$ can be written in terms of generally more accessible quantities as $kw/\alpha\eta$, where $kw$ is the fault-zone hydraulic transmissivity, $\eta$ is the fluid dynamic viscosity, and $\alpha$ is the fault-zone hydraulic diffusivity. At the centimetric scale composing the smallest aseismic slip events in the dataset, Passelègue \textit{et al.} \cite{Passelegue_Almakari_2020} estimated the hydraulic transmissivity of their saw-cut granitic fault within 10$^{-17}$ and 2$\times$10$^{-18}$ m$^3$, and a hydraulic diffusivity from 3$\times$10$^{-5}$ m$^2$/s to 10$^{-6}$ m$^2$/s \cite{Almakari_Chauris_2020}, at confining pressures ranging from 20 to 100 MPa respectively. Considering a water dynamic viscosity at the room temperature the experiments were conducted, $\eta\sim10^{-3}$ Pa$\cdot$s, we estimate $wS$ to be within 3$\times$10$^{-10}$ and 2$\times$10$^{-9}$ m/Pa (assuming that $kw$ and $\alpha$ are positively correlated). Taking into consideration the aforementioned characteristic range of values for $f_r$ and $\Delta \tau_{r-0}$, we estimate the maximum value for the in-situ factor that is representative of these laboratory experiments to be roughly $I_\text{situ}\sim10^{12}$ N$\cdot$m$^{-7/2}$. Interestingly, the upper bound for the moment release resulting from this value of $I_\text{situ}$ aligns closely with our estimates of moment release and injected fluid volumes for this very same set of experiments (Fig. \ref{fig:moment-observations}, yellow triangles). Note that in Fig. \ref{fig:moment-observations}, the factor $N$ must always be interpreted as being greater than $I_\text{situ}$ due to the combined effect of the coefficients $C_\nu$ and $C_\text{shut-in}$. In addition, this upper bound seems to explain relatively well the centimeter-scale laboratory experiments presented in this study (cyan triangles; Supplementary Materials) and the 
meter-scale laboratory experiments of Cebry \textit{et al.} \cite{Cebry_Ke_2022} (red triangles). The former experiments were carried out under almost identical conditions to the ones of Passelègue \textit{et al.} \cite{Passelegue_Almakari_2020}, whereas the latter ones were conducted in a similar saw-cut granitic fault with hydraulic properties that are close to the ones of Passelègue \textit{et al.}'s fault at the lower confining pressures of this latter one \cite{Passelegue_Almakari_2020}.

At the large scale of industrial fluid injections, we consider one of the best-documented field cases: the 1993 hydraulic stimulation at the Soultz geothermal site in France \cite{Cornet_Helm_1997}. The hydraulic transmissivity associated with the 550-m open-hole section stimulated during the test has been estimated to experience a 200-fold increase as a consequence of the two fluid injections conducted, giving us a possible range of approximately 10$^{-14}$ m$^3$ to 2$\times$10$^{-12}$ m$^3$ \cite{Evans_Genter_2005}. However, the smallest value of $kw$ represents only the very short, initial part of the injection \cite{Evans_Genter_2005}. Therefore, a possible variation between 5$\times$10$^{-14}$ m$^3$ and 2$\times$10$^{-12}$ m$^3$ seems a more reasonable range to be considered within the assumptions of our model which assumes a constant transmissivity. On the other hand, the hydraulic diffusivity possesses significant uncertainties due to the single-well nature of the hydraulic data in contrast to the double-well measurements employed, for instance, by Passelègue \textit{et al.} \cite{Passelegue_Almakari_2020} in the laboratory. We consider a range of values for $\alpha$ from 0.01 m$^2$/s to 0.1 m$^2$/s, which is consistent with estimates derived from micro-seismicity migration \cite{Parotidis_Shapiro_2004} and aseismic fracture slip \cite{Saez_Lecampion_2023}. Assuming a water dynamic viscosity of $\eta=2\times10^{-4}$ Pa$\cdot$s which is representative of the temperature conditions within the reservoir \cite{Evans_Moriya_2005}
, we estimate $wS$ to fall within the range of $5\times10^{-8}$ m/Pa to $10^{-7}$ m/Pa. With these estimates, we calculate a representative maximum value for the in-situ factor in this field test to be roughly $I_\text{situ}\sim10^{9}$ N$\cdot$ m$^{-7/2}$. As shown in Fig. \ref{fig:moment-observations}, the resulting upper limit aligns very well with the field data (circles), providing an effective upper bound for the hectometric to kilometric rupture cases composing the dataset. Furthermore, this simplified, order-of-magnitude analysis suggests that the behavior of the upper limit we observe from the laboratory to the reservoir scale, namely, the decrease of the factor $N$ with increasingly larger volume and moment release scales, might be primarily controlled by an increase in fault-zone storativity. 
Moreover, the upper bound for intermediate scales (in-situ experiments, square symbols in Fig. \ref{fig:moment-observations}) is characterized by a value of $N$ (or $I_\text{situ}$) that is approximately in the middle of the values that provide an upper limit for the laboratory and field data, suggesting that the increase in storativity with larger scales could be a general explanation for the trend observed throughout the entire dataset.

Finally, Fig.~\ref{fig:size-observations} shows the comparison of our scaling relation for the maximum rupture run-out distance, equation \eqref{eq:Rmax-nearly-unstable}, with the estimated rupture run-out distances for the same injection-induced aseismic slip events as in Fig. \ref{fig:moment-observations}. To facilitate the interpretation, we similarly define the factor $M=S_\nu A_\text{situ}$ accounting for all effects other than the injected fluid volume, so that equation \eqref{eq:Rmax-nearly-unstable} becomes simply $R_\text{max}=M\sqrt{V_\text{tot}}$. It is important to note that the effects of $\Delta\tau_{r-0}$ and $wS$ are now of similar order, as $R_\text{max}$ scales alike with the background stress change $R_\text{max}\propto \Delta\tau_{r-0}^{-1/2}$ and fault-zone storativity $R_\text{max}\propto (wS)^{-1/2}$. Considering the same range of values for $f_r$, $\Delta\tau_{r-0}$ and $wS$ discussed previously, we calculate $A_\text{situ}$ to be around 200 m$^{-1/2}$ for Passelègue \textit{et al.}'s experiments, and 15 m$^{-1/2}$ for the Soultz case. As shown in Fig. \ref{fig:size-observations}, the upper limits resulting from these two values of $A_\text{situ}$ (considering also an amplification due to the factor $S_\nu$) are in remarkably good agreement with the data, providing an effective upper bound for the maximum rupture run-out distance from cm-scale ruptures in the laboratory to km-scale ruptures in industrial applications.

\section{Discussion}

Our results provide a rupture-mechanics-based estimate for the maximum size, moment release, and magnitude of injection-induced slow slip events. Moreover, the dependence of our scaling relations on in-situ conditions and injected fluid volume allows us to explain variations in rupture sizes and moment releases resulting from fluid injections that span more than 12 orders of magnitude of injected fluid volume. 
Similar scaling relations for the moment release of regular, fast earthquakes have been previously proposed in the literature \cite{McGarr_2014,Galis_Ampuero_2017,Li_Elsworth_2021}. Notably absent in those relations, fault-zone storativity appears here to be a crucial factor influencing the upper-bound behavior we observe with increasing scales of volume and moment release in the data. While our scaling relation for the maximum aseismic rupture run-out distance is the first of its kind, McGarr and Barbour \cite{McGarr_Barbour_2018} suggested in a prior work that for the moment release, the relation for the cumulative moment $\Sigma M_0=2\mu V_\text{tot}$ that was originally proposed by McGarr \cite{McGarr_2014} for regular earthquakes, account also for aseismic slips. It is thus pertinent to discuss their scaling relation in light of our findings.

We first note that in testing their relation, McGarr and Barbour \cite{McGarr_Barbour_2018} incorporated numerous data points of aseismic moment release and injected volume into a dataset characterized by otherwise only regular earthquakes. All of these aseismic slip events come from laboratory experiments of hydraulic fracturing \cite{Goodfellow_Nasseri_2015}, except for one single data point that stems from direct measurements of injection-induced aseismic slip during an in-situ experiment \cite{Guglielmi_Cappa_2015}. The mechanics of hydraulic fractures \cite{Detournay_2016}, however, differs significantly from its shear rupture counterpart. Indeed, the moment release by hydraulic fractures scales linearly with the injected fluid volume simply because the integral of the fracture width over the crack area is equal to the fracture volume. The latter is approximately equal to the injected volume under common field conditions, namely, negligible fluid leak-off and fluid lag \cite{Detournay_2016}. In our study, we discarded these hydraulic-fracturing data points because they correspond to a different phenomenon. The remaining data point of McGarr and Barbour, which does correspond to a fluid-driven shear rupture \cite{Guglielmi_Cappa_2015}, is retained in our dataset albeit with a certain degree of uncertainty based on moment release estimates provided by more recent studies (Supplementary Materials).

In terms of modeling assumptions, one of the most significant differences between McGarr's and ours is that we account for the potential for aseismic ruptures to propagate beyond the fluid-pressurized region ($\lambda\gg1$). This regime, which from an upper-bound perspective is of most practical interest as it produces the largest ruptures for a given injection, is not allowed by construction in McGarr's model due to his assumption that any fault slip induced by the fluid injection must be confined within the region where pore fluids have been effectively pressurized due to the injection \cite{McGarr_2014}. We emphasize that aseismic ruptures breaking non-pressurized fault regions are a possibility that always emerges when incorporating rupture physics in a model \cite{Garagash_Germanovich_2012}, even in the absence of frictional weakening owing simply to long-range elastostatic stress transfer effects \cite{Bhattacharya_Viesca_2019}. Moreover, such a regime has already been directly observed in laboratory experiments \cite{Cebry_Ke_2022}, and inferred to have occurred during in-situ experiments \cite{Guglielmi_Cappa_2015,Bhattacharya_Viesca_2019} and industrial fluid injections for reservoir stimulation \cite{Eyre_Eaton_2019}. Furthermore, as a natural consequence of incorporating rupture physics in our model, we obtain a dependence of the moment release on the background stress state and fault frictional parameters. McGarr's model is, in contrast, insensitive to these physical quantities, which largely control the release of elastic strain energy during rupture propagation. Another important distinction between both models is that McGarr's relies uniquely on the capacity of the rock bulk to elastically deform and volumetrically shrink to accommodate the influx of fluid mass from the injection, unlike our model which accounts for bulk, fluid, and pore compressibilities within the fault zone via the so-called oedometric storage coefficient \cite{Green_Wang_1990}.

Despite the significant differences between both models, it is pertinent to compare McGarr's relation for the moment release with our newly compiled dataset. By doing so, we observe that McGarr's upper bound can explain the majority of the data points, albeit with one very important exception (Fig. \ref{fig:moment-observations}): the 2017 $M_w$ 5.0 slow slip event in western Canada \cite{Eyre_Samsonov_2022}; the largest event detected thus far. Specifically, McGarr's formula fails by predicting a maximum magnitude of 4.4 (considering $V_\text{tot}=88,473$ m$^3$ and assuming a shear modulus of 30 GPa \cite{Eyre_Samsonov_2022}). This magnitude is equivalent to predicting an upper limit for the moment release that is 16 times smaller than the actual moment that was inferred geodetically \cite{Eyre_Samsonov_2022}. Such underestimation is somewhat similar to that performed by McGarr's formula in the case of regular earthquakes: for instance, when considering the 2017 $M_w$ 5.5 Pohang earthquake in South Korea \cite{Ellsworth_Giardini_2019}. Our scaling relation can, conversely, explain the $M_w$ 5.0 slow slip event in Canada and, more generally, our entire compilation of events by accounting for variations in in-situ conditions such as the background stress change ($\Delta \tau_{r-0}$) and, especially, the fault-zone storativity ($wS$). Note that from a `data-fitting' perspective, the dependence of our model on $wS$ and $\Delta \tau_{r-0}$ introduces additional degrees of freedom compared to McGarr's formula, which depends only on the injected fluid volume and the shear modulus; a parameter that has very little variation in practice.

Our estimates of the maximum rupture size and magnitude for slow slip events may be regarded, to some extent, as an aseismic counterpart of the also rupture-mechanics-based scaling relation proposed by Galis \textit{et al.} \cite{Galis_Ampuero_2017} for regular earthquakes. Although we are describing a fundamentally different process here, the two scaling relations share the same 3/2-power law dependence on the injected fluid volume. This equal exponent arises from the similarities between the competing forces driving both slow slip events and dynamic ruptures in each model, namely, a point-force load due to fluid injection and a uniform stress change behind the cohesive zone and within the ruptured surface. In Galis \textit{et al.}'s model, a point-force-like load is imposed to nucleate an earthquake. In our model, it is the natural asymptotic form that the equivalent force associated with the fluid injection takes in the regime that provides the largest ruptures for a given injection ($\lambda\gg1$). Note that the two models differ in their storativity-like quantity. As we discussed before, we account for the capacity of the fluid, pore space, and bulk material in the fault zone to store pressurized fluids. In contrast, Galis \textit{et al.}'s model accounts only for the capacity of the bulk material: a property they inherited from McGarr's model \cite{McGarr_2014}. A revision of seismic scaling relations may be required to include the notion of a more general storativity term, particularly considering the significant variability in pore compressibility observed in practice \cite{Kuang_Jiao_2020,Doan_Brodsky_2006,Xue_Brodsky_2013,Rutqvist_Noorishad_1998} which can sometimes dominate over bulk and fluid compressibilities. Another important difference with Galis \textit{et al.}'s model is the uniform variation of background stress which in their model is the so-called stress drop $\Delta\tau_{0-r}=\tau_0-f_r\sigma_0^\prime$, whereas in our case, it corresponds to the same quantity but of opposite sign, $\Delta\tau_{r-0}=f_r\sigma_0^\prime-\tau_0$. Conceptually, this is indeed a very important difference. In our model, the stress \textit{drop} is negative, which implies that after the termination of the injection operation and the subsequent dissipation of overpressure due to the injection, the residual shear stress acting on the slipped fault patch will be greater than the initial shear stress. This, which is a prominent feature of unconditionally stable ruptures, implies no release of tectonically accumulated pre-stresses on the fault.

The previous point brings us to an important issue: we have considered only one of the two possible modes of aseismic slip, namely, fault ruptures that are unconditionally stable. However, injection-induced aseismic slip can also be the result of conditionally stable slip, that is, the nucleation phase preceding an otherwise dynamic rupture. The principal factor determining whether aseismic slip will develop in one way or the other is the so-called ultimate stability condition \cite{Garagash_Germanovich_2012,Saez_Lecampion_2024}. For conditionally stable slip to occur, the initial shear stress must be thus greater than the background residual fault strength ($\tau_0>f_r\sigma_0^\prime$), resulting in a positive stress drop. This is therefore the mode of aseismic slip that can potentially release tectonically accumulated pre-stresses. In general, we cannot rule out that the points in the datasets of Fig. \ref{fig:moment-observations} and \ref{fig:size-observations} correspond to either conditionally stable or unconditionally stable slip, as estimating the background stress state and fault frictional properties that are representative of the reactivated fault remains extremely challenging in practice. There is, nevertheless, at least one case in the dataset in which aseismic slip is as a matter of fact, conditionally stable. These are the two aseismic slip events from the meter-scale laboratory experiments of Cebry \textit{et al.} \cite{Cebry_Ke_2022}, which preceded seismic ruptures that broke the entire fault interface sample (Supplementary Materials). The scaling relations resulting from this mode of aseismic slip are therefore important. Moreover, considering that the data points from Cebry \textit{et al.}'s experiments align well with the other points in the dataset, we anticipate these scaling relations to be similar in their structure to the ones presented here.

Our model aimed to capture the most essential physical ingredients of unconditionally stable ruptures to provide the desired theoretical insights into the physical mechanisms controlling the maximum size and magnitude of injection-induced slow slip events. To achieve this, we have however adopted several simplifying assumptions that warrant further investigation. In particular, our model does not account for fluid leak-off from the permeable fault zone to the host rock, nor permeability enhancements associated with fault slip and/or the reduction of effective normal stress due to fluid injection. Despite these simplifications, we expect our scaling relations to still provide an effective upper bound with regard to these additional factors. Indeed, we think that incorporating a permeable host rock would notably decrease the injection overpressure in the fault zone compared to the impermeable case, thus decelerating rupture growth. The effect of slip-induced dilatancy, which is relatively well-established \cite{Ciardo_Lecampion_2019,Dunham_2024}, would introduce a toughening effect that would similarly slow down slip propagation from a fracture-mechanics perspective. Furthermore, permeability enhancements due to both dilatancy and reduced effective normal stress are expected to be inconsequential in the limit $\lambda \gg 1$, which is the relevant one for establishing an upper bound. This is due to, in this regime, most of the slipping region remains non-pressurized except for a small area near the fluid source. The strength of this small (point-force-like) region remains unchanged in our model, provided that the enhanced hydraulic properties are considered as the constant ones \cite{Dunham_2024}. An additional simplification in our model is the consideration of a single fault zone. Although this might likely be the case for the majority of the events incorporated in our dataset \cite{Passelegue_Almakari_2020,Cebry_Ke_2022,Guglielmi_Cappa_2015,DeBarros_Daniel_2016,Duboeuf_DeBarros_2017,Lengline_Boubacar_2017,Eyre_Eaton_2019,Eyre_Samsonov_2022}, in some cases, a network of fractures or faults could be reactivated instead \cite{Cornet_Helm_1997,Evans_Moriya_2005}. Recent numerical modeling studies on injection-induced aseismic slip have, however, shown that approximately the same scaling relations for the moment release predicted by a single fracture in two dimensions emerge collectively for a set of reactivated fractures belonging to a two-dimensional discrete fracture network \cite{Ciardo_Lecampion_2023}. This is notably the case when the regime $\lambda\gg1$ is reached in a global, fracture-network sense. Yet the generality and prevalence of this finding in three dimensions remain to be confirmed.

We notably showed that in the nearly unstable regime ($\lambda\gg1$), the dynamics of the rupture expansion in our upper-bound configuration are controlled uniquely by the history of injected fluid volume, irrespective of any other characteristic of the injection protocol. The implications of this finding may go well beyond the ones explored in this work. For example, in hydraulic stimulation operations for the development of deep geothermal energy, micro-seismicity clouds which often accompany fluid injections are commonly used to constrain the areas of the reservoir that have been effectively stimulated. If aseismic-slip stress transfer is a dominant mechanism in the triggering of micro-seismicity, our model suggests that these seismicity clouds may contain important information about the pre-injection stress state and fault frictional properties which are embedded in the factor $A_\text{situ}$ (equation \eqref{eq:R-volume-Q-const}). Moreover, if the effect of the fracture energy on rupture propagation can be approximately neglected in comparison to the other two competing forces driving aseismic ruptures in our model, that is, the background stress change and fluid-injection force, our results imply that the spatiotemporal patterns of seismicity migration might be deeply connected to injection protocols via the dependence of the aseismic slip front dynamics on the square root of the cumulative injected fluid volume. This could be used, for instance, to identify from injection-induced seismicity catalogs under what conditions aseismic-slip stress transfer may become a potentially dominant triggering mechanism due to this unique spatiotemporal footprint, which differs notably from the ones emerging from other triggering mechanisms such as pore pressure diffusion and poroelastic stressing \cite{Shapiro_2015,Goebel_Brodsky_2018}. Similarly, our model could be potentially applied to the study of natural seismic swarms where sometimes fluid flow and aseismic slip processes are thought to be the driving forces behind their observed dynamics \cite{Sirorattanakul_Ross_2022,Yukutake_Yoshida_2022}. Lastly, our model could be also utilized to understand slow slip events occurring at tectonic plate boundaries in many subduction zones worldwide. The fundamental mechanics of slow slip events remains debated \cite{Burgmann_2018} yet multiple, recent observations suggest that their onset and arrest might be spatially and temporally correlated with transients of pore-fluid pressure \cite{Tanaka_Kato_2010,Nakajima_Uchida_2018,Warren-Smith_Fry_2019}.

Finally, we emphasize that our investigation has focused on constraining the rupture size and moment release of purely aseismic injection-induced ruptures. However, in some instances, seismic or micro-seismic events may release a substantial portion of the elastic strain energy stored in the medium. In this study, we have incorporated in our compilation of events only cases where the seismic contribution to the moment release is thought to be orders of magnitude smaller than the aseismic part. From a mechanics perspective, this aimed to exclude events where the stress transfer from frictional instabilities could significantly influence the dynamics of the slow rupture under consideration, thereby ensuring a robust comparison between the data and the scaling relations of our model. Future studies should therefore focus on understanding what physical factors govern the partitioning between aseismic and seismic slips during injection operations. Our work, in this sense, contributes to such possibility by providing an upper limit to the previously unexplored aseismic end-member. Together with prior works on purely seismic ruptures, we believe this offers a starting point to examine slip partitioning during injection-induced fault slip sequences: a crucial step toward advancing our physical understanding of the seismogenic behavior of reactivated faults and the associated seismic hazard.

\section{Materials and Methods}\label{sec:methods}

\subsection{Time-dependent upper-bound model for the size of unconditionally stable ruptures} \label{subsec:time-dependent-upper-bound-model}

We define the axisymmetric overpressure due to the injection as $\Delta p(r,t)=p(r,t)-p_0$, with $p_0$ the uniform background pore pressure. During the pressurization stage ($t\leq t_s$), the overpressure is given by $\Delta p(r,t)=\Delta p_*\cdot E_1\left(r^2/4\alpha t\right)$ for an injection at constant flow rate $Q_0$ \cite{Carslaw_Jaeger_1959}, where $\Delta p_*=Q_0\eta/4\pi kw$ is the intensity of the injection with units of pressure, $\alpha$ is the fault hydraulic diffusivity, $\eta$ is the fluid dynamic viscosity, the product $kw$ is the so-called fault hydraulic transmissivity, and $E_1$ is the exponential integral function. The fracture-mechanics energy balance for a quasi-static circular rupture propagating on a slip-weakening fault was presented in \cite{Saez_Lecampion_2024}. In our upper-bound configuration here, neglecting the fracture energy spent during rupture propagation leads to the following expression describing the evolution of the rupture radius $R$ with time:
\begin{equation}\label{eq:energy-balance}
    \frac{2}{\sqrt{\pi}}\frac{f_{r}\Delta p_{*}}{\sqrt{R(t)}}\int_{0}^{R(t)}\frac{E_1\left(r^2/4\alpha t\right)}{\sqrt{R(t)^2-r^2}}r\textrm{d}r =
    \frac{2}{\sqrt{\pi}}\Delta\tau_{r-0}\sqrt{R(t)},
\end{equation}
where $\Delta\tau_{r-0}=f_{r}\sigma_{0}^{\prime}-\tau_{0}$ is the 
background stress change. In equation \eqref{eq:energy-balance}, the integral term of the left-hand side 
is associated with an influx of potential energy towards the rupture front which becomes available for the rupture to grow owing to the sole effect of overpressure due to the injection. Conversely, the term of the right-hand side due to the background stress change is responsible alone for resisting rupture advancement. Nondimensionalization of equation \eqref{eq:energy-balance} shows that the competition between both \textit{energy} terms is quantified by one single dimensionless number, the so-called residual stress-injection parameter $\mathcal{T}_r= \Delta\tau_{r-0}/f_r\Delta p_*$, introduced first in \cite{Saez_Lecampion_2024}. Moreover, equation \eqref{eq:energy-balance} admits analytical solution in the form: $R(t)=\lambda \cdot L(t)$ \cite{Saez_Lecampion_2022}, with the asymptotes $\lambda \simeq 1/\sqrt{2\mathcal{T}_r}$ for nearly unstable ruptures ($\lambda\gg 1, \mathcal{T}_r \ll1$), and $\lambda \simeq e^{(2-\gamma-\mathcal{T}_r)/2}/2$ for marginally pressurized ruptures ($\lambda\ll 1, \mathcal{T}_r \sim 10 $). 
To highlight how significant is to analyze the end-member cases of nearly unstable ($\lambda\gg1$) and marginally pressurized ($\lambda\ll1$) ruptures throughout this work, we refer to their asymptotes for $\lambda$ plotted in Fig. \ref{fig:upperbound-lambda}b which nearly overlap and thus quantify together almost any rupture scenario. 
This analytical solution for $\lambda$ was first derived by Sáez \textit{et al.} \cite{Saez_Lecampion_2022} for a fault interface with a constant friction coefficient (equation (21) in \cite{Saez_Lecampion_2022}). Here, in our upper-bound configuration, the mathematical solution is identical to the one presented in \cite{Saez_Lecampion_2022} provided that the constant friction coefficient $f$ in \cite{Saez_Lecampion_2022} is understood as the residual value $f_r$ of the slip-weakening friction law here.

To make the link between the evolution of the rupture radius $R(t)$ and the injected fluid volume $V(t)$ in the most practically relevant, nearly unstable regime ($\lambda\gg1$), we use the asymptote $R(t)\simeq(1/\sqrt{2\mathcal{T}_r})L(t)$ in combination with the following expressions for the residual stress-injection parameter $\mathcal{T}_r=\Delta\tau_{r-0}/f_r\Delta p_*$, overpressure intensity $\Delta p_*=Q_0\eta/4\pi kw$, overpressure front $L(t)=\sqrt{4\alpha t}$, hydraulic diffusivity $\alpha=k/S\eta$, and injected fluid volume $V(t)=Q_0 t$. By doing so, we arrive at equation \eqref{eq:R-volume-Q-const} in the main text. For non-circular ruptures ($\nu\neq0$), building upon the work of Sáez \textit{et al.} \cite{Saez_Lecampion_2022} for a constant friction coefficient, we obtain that the rupture front of our upper-bound model is well-approximated by an elliptical shape that becomes more elongated for increasing values of $\nu$ and decreasing values of $\mathcal{T}_r$, with a maximum aspect ratio of $1/(1-\nu)$ when $\mathcal{T}_r \ll 1$ and a minimum aspect ratio of $(3-\nu)/(3-2\nu)$ when $\mathcal{T}_r \sim 10$. Other features of Sáez \textit{et al.}'s model \cite{Saez_Lecampion_2022} such as the invariance of the rupture area with regard to the Poisson's ratio and the numerically-derived asymptotes for the quasi-elliptical fronts are also inherited here in the upper-bound model. In the shut-in stage ($t>t_s$), the overpressure is obtained by superposition simply as $\Delta p(r,t)=\Delta p_*\cdot \left[ E_1\left(r^2/4\alpha t\right) - E_1\left(r^2/4\alpha (t-t_s)\right) \right]$. The spatiotemporal evolution of overpressure has been studied in detail in \cite{Saez_Lecampion_2023}. Moreover, as already discussed in the main text, we reduced the upper-bound problem in the shut-in stage to a fault responding with a constant friction coefficient equal to $f_r$. Hence, our upper-bound model inherits all the results obtained by Sáez and Lecampion \cite{Saez_Lecampion_2023} who investigated extensively the propagation and arrest of post-injection aseismic slip on a fault obeying a constant friction coefficient. In particular, we take advantage of their understanding of the propagation and arrest of the slip front that ultimately determines the maximum size of unconditionally stable ruptures in our upper-bound model. Here, we have indeed expanded the work of Sáez and Lecampion \cite{Saez_Lecampion_2023} to account for an examination of the previously unknown evolution of the moment release during the shut-in stage (Fig. \ref{fig:maximum-moment}).

\subsection{Asymptotics of moment release for nearly unstable and marginally pressurized ruptures} \label{subsec:slip-profiles}

The scalar moment release $M_0$ at a given time $t$ is given by \cite{Aki_2002}, 
\begin{equation}\label{eq:moment-release-definition}
    M_0(t)=\mu\iint_{A_r(t)}\delta(x,y,t)\text{d}x\text{d}y,
\end{equation}
where $\mu$ is the bulk shear modulus, $\delta$ is the current slip distribution, and $A_r$ is the current rupture surface. To calculate the time-dependent slip distribution in the circular rupture case, we consider the quasi-static relation between fault slip $\delta$ and the associated elastic change of shear stress $\Delta\tau$ within an axisymmetric circular shear crack \cite{Sneddon_1951}:
\begin{equation}\label{eq:elastic-equilibrium-circular-inverse}
    \delta(r,t)=\frac{4R(t)}{\pi\mu}\int_{\bar{r}}^{1}\frac{\xi\mathrm{d}\xi}{\sqrt{\xi^{2}-\bar{r}^{2}}}\int_{0}^{1}\frac{\Delta\tau(s\xi R(t),t)s\mathrm{d}s}{\sqrt{1-s^{2}}},
\end{equation}
where $\bar{r}=r/R(t)$ is the normalized radial coordinate. Equation \eqref{eq:elastic-equilibrium-circular-inverse} was originally derived for an internally-pressurized tensile circular crack with axisymmetric load \cite{Sneddon_1951}. Nevertheless, under the assumptions of uni-directional slip with axisymmetric magnitude and a Poisson's ratio $\nu=0$, the shear crack problem is mathematically equivalent on the fault plane to its tensile counterpart \cite{Bhattacharya_Viesca_2019}: crack opening being $\delta$ and crack-normal stress change being $\Delta\tau$. In the limiting regime of a rupture propagating with zero fracture energy and at the residual friction level $f_r$, the change of shear stress is simply 
\begin{equation}\label{eq:stress-change}
    \Delta\tau(r,t)=\tau_0-f_r\left[\sigma_0^{\prime}-\Delta p(r,t)\right]=f_r\Delta p(r,t)-\Delta \tau_{r-0},
\end{equation}
where $\Delta \tau_{r-0}=f_r\sigma_0^\prime-\tau_0$ is the background stress change. Hence, for injection at a constant volumetric rate $Q_0$, the spatio-temporal evolution of slip for the end-member cases of nearly unstable ($\lambda\gg1$) and marginally pressurized ($\lambda\ll1$) ruptures turn out to be identical to the ones determined by Sáez \textit{et al.} \cite{Saez_Lecampion_2022} for their so-called critically stressed regime (equation (26) in \cite{Saez_Lecampion_2022}) and marginally pressurized regime (equation (25) in \cite{Saez_Lecampion_2022}) respectively, as long as we interpret their constant friction coefficient $f$ as $f_r$. The self-similar slip profiles can be written in a more convenient dimensionless form as: 
\begin{equation}\label{eq:self-similar-slip-profiles}
    \delta(r,t)/\delta_*(t)=D\left(r/R(t)\right),\quad\text{with}\quad
    \delta_*(t)=
    \begin{cases}
        \Delta \tau_{r-0}R(t)/\mu & \text{when }\lambda\gg1,\\
        f_r\Delta p_* R(t)/\mu & \text{when } \lambda\ll1,
    \end{cases}
\end{equation}
and
\begin{equation}
    D(x)=
    \begin{cases}
        (4/\pi)(\text{arccos}(x)/x-\sqrt{1-x^2}) & \text{when }\lambda\gg1,\\
        (8/\pi)(\sqrt{1-x^2}-x\cdot\text{arccos}(x)) & \text{when } \lambda\ll1.
    \end{cases}
\end{equation}
Note that in the nearly unstable regime ($\lambda\gg1$), we have recast equation (26) in \cite{Saez_Lecampion_2022} using the expressions $L(t)=R(t)/\lambda$ and $\lambda\simeq1/\sqrt{2\mathcal{T}_r}$. The nearly unstable asymptote for fault slip is plotted in Fig. \ref{fig:maximum-size}b and compared to the numerical solution. Integration of the self-similar slip profiles via equation \eqref{eq:moment-release-definition} leads to the asymptotes for the moment release during the pressurization stage given in the main text: $M_0(t)=(16/3)\Delta \tau_{r-0}R(t)^3$ when $\lambda\gg1$, and $M_0(t)=(16/9)f_r\Delta p_* R(t)^3$ when $\lambda\ll1$. It is worth mentioning that the slip distribution of nearly unstable ruptures has a singularity (of order $1/r$) at $r=0$. Strictly speaking, this asymptote corresponds to the solution of the so-called outer problem which is defined at distances $r\gg L(t)$. An interior layer must be resolved at distances $r\sim L(t)$ to obtain the finite slip at the injection point, which scales as $\delta_c(t)=f_r\Delta p_* L(t)/\mu$ \cite{Saez_Lecampion_2022} (see Fig. \ref{fig:maximum-size}b). Nevertheless, this interior layer has no consequences in estimating the moment release. Indeed, the integrand in equation \eqref{eq:moment-release-definition} for such a slip distribution is non-singular so that after taking the limit $L(t)/R(t)\to0$, one effectively recovers the actual asymptote for the moment release. The details of the interior layer are therefore irrelevant to the calculation of $M_0$ in this limit.

\subsection{Relation between fluid-injection force and injected fluid volume for arbitrary fluid sources}\label{subsec:volume-relation}

Under the assumptions of our model, the displacement field $\boldsymbol{u}$ induced by the fluid injection into the poroelastic fault zone is irrotational $\nabla \times \boldsymbol{u}=0$ \cite{Marck_Savitski_2015}. Therefore, the variation in fluid content $\zeta$, which corresponds to the change of fluid volume per unit volume of porous material with respect to an initial state (here, $t=0$), satisfies the following constitutive relation with the pore-fluid overpressure $\Delta p$ (eq. 96, \cite{Detournay_Cheng_1993}),
\begin{equation}\label{eq:fluid-content-increase-pressure}
    \zeta=S \Delta p,
\end{equation}
where $S$ is the so-called oedometric storage coefficient representing the variation of fluid content caused by a unit pore pressure change under uniaxial strain and constant normal stress in the direction of the strain \cite{Green_Wang_1990}, here, the $z$-axis (Fig. \ref{fig:model-schematics}c). 
$S$ accounts for the effects of fluid, pore, and bulk compressibilities of the fault zone, and is equal to \cite{Detournay_Cheng_1993}
\begin{equation}\label{eq:storage-coefficient-definition}
    S=\frac{1}{M}+\frac{b^2(1-2\nu)}{2(1-\nu)\mu},
\end{equation}
where $M$ is the Biot's modulus and $b$ the Biot's coefficient.

To obtain the cumulative injected fluid volume at a given time, $V(t)$, we just sum up changes in fluid volume all over the spatial domain of interest, say $\Omega$, at a given time $t$, that is, $V(t)=\int_\Omega \zeta \text{d}\Omega$. In our model, the fluid flow problem is axisymmetric and the fault-zone width $w$ is uniform, such that the differential of the volume is simply $\text{d}\Omega=2\pi w r\text{d}r$ in cylindrical coordinates. With these definitions, we can now integrate \eqref{eq:fluid-content-increase-pressure} over the entire fault-zone volume to obtain the following expression for the injected fluid volume valid for an arbitrary fluid injection: 
\begin{equation}\label{eq:volume-integral}
    V(t)=S w \cdot 2\pi \int_0^\infty \Delta p(r,t)r\text{d}r.
\end{equation}

By defining the normal force induced by the fluid injection over the slip surface (simply equal to the integral of the overpressure over the fault plane) as:
\begin{equation}\label{eq:fluid-injection-force}
    F(t)=2\pi\int_0^\infty \Delta p(r,t)r\text{d}r,
\end{equation}
we arrive at the following relation between the fluid-injection force and injected fluid volume:
\begin{equation}\label{eq:point-force-volume}
    F(t)=\frac{V(t)}{wS}.
\end{equation}
Expressions of a similar kind to \eqref{eq:point-force-volume} have been reported in previous studies \cite{Shapiro_Dinske_2010,McGarr_2014,Garagash_2021}. For example, McGarr \cite{McGarr_2014} considered a similar relation except that his storativity-like term is the inverse of the elastic bulk modulus. Garagash \cite{Garagash_2021} also proposed a similar expression to \eqref{eq:point-force-volume} but accounting only for pore compressibility. Finally, the relation considered by Shapiro \textit{et al.} \cite{Shapiro_Dinske_2010} is the closest to our expression, including the oedometric storage coefficient.

\subsection{Scaling relations for nearly unstable ruptures accounting for arbitrary fluid injections}\label{sec:generalization}

Nearly unstable ruptures ($\lambda\gg1$) provide the upper bound of most practical interest. Here, we generalize such an upper bound for the rupture size and moment release to account for an arbitrary fluid injection. In the pressurization stage ($t\leq t_s$), the reduction of fault strength due to fluid injection in the so-called outer problem ($r\gg L(t)$) can be effectively approximated as a point force (e.g., \cite{Garagash_Germanovich_2012,Saez_Lecampion_2022}),
\begin{equation}\label{eq:point-force-definition}
    f_r\Delta p(r,t) \approx f_r F(t)\frac{\delta^{\text{dirac}}(r)}{2\pi r}= f_r \frac{V(t)}{wS}\frac{\delta^{\text{dirac}}(r)}{2\pi r},
\end{equation}
where $F(t)$ is the fluid-injection normal force, equation \eqref{eq:fluid-injection-force}, which is related to the cumulative injected fluid volume via equation \eqref{eq:point-force-volume}. Substituting equation \eqref{eq:point-force-definition} into the stress change \eqref{eq:stress-change}, and then the latter into the double integral, equation \eqref{eq:elastic-equilibrium-circular-inverse}, we obtain upon evaluating those integrals an asymptotic upper bound for the spatiotemporal evolution of fault slip  as:
\begin{equation}\label{eq:slip-distribution-with-singularity}
    \delta(r,t)=\frac{4}{\pi}\frac{\Delta \tau_{r-0}}{\mu}R(t)\left[ \frac{f_rV(t)}{2\pi w S\Delta\tau_{r-0}}\frac{\text{arccos}\left(r/R(t)\right)}{r/R(t)} - \sqrt{1-\left(r/R(t)\right)^2} \right],
\end{equation}

The propagation condition for a rupture with negligible fracture energy, equation \eqref{eq:energy-balance}, can be alternatively written in terms of the slip behavior near the rupture front as \cite{Barenblatt_1962},
\begin{equation}\label{eq:prop-condition-on-slip}
    \lim_{r\to R(t)^-}\frac{\partial\delta(r,t)}{\partial r}\sqrt{R(t)-r}=0.
\end{equation}
This imposes a constraint in the slip distribution \eqref{eq:slip-distribution-with-singularity} that can be also seen, in a limiting sense, as eliminating any stress singularity at the rupture front. By differentiating equation \eqref{eq:slip-distribution-with-singularity} with respect to $r$, and then applying the propagation condition \eqref{eq:prop-condition-on-slip}, we obtain the following relation:
\begin{equation}\label{eq:rupture-front-volume-relation}
    R(t)=\sqrt{\frac{f_r V(t)}{2\pi wS \Delta\tau_{r-0}}},
\end{equation}
which is valid for an arbitrary fluid injection. 

Equation \eqref{eq:rupture-front-volume-relation} is identical to equation \eqref{eq:R-volume-Q-const} in the main text, which was originally derived for injection at constant flow rate. It thus represents a generalization of the insightful relation between the evolution of the rupture radius and the cumulative injected fluid volume, equation \eqref{eq:R-volume-Q-const}, for arbitrary fluid injections. Note that alternatively, equation \eqref{eq:rupture-front-volume-relation} can be derived through the rupture propagation condition imposed over the stress change, equation \eqref{eq:energy-balance}. It only takes to replace the particular overpressure solution for injection at a constant volumetric rate, $\Delta p(r,t)=\Delta p_*E_1\left(r^2/4\alpha t\right)$, by the more general point-force representation, equation \eqref{eq:point-force-definition}. We report here the derivation based on the slip distribution because it makes now the calculation of the moment release for arbitrary fluid injections straightforward. Indeed, by substituting \eqref{eq:rupture-front-volume-relation} into \eqref{eq:slip-distribution-with-singularity}, we obtain the slip distribution satisfying the zero-fracture-energy condition of our upper-bound model. Upon integrating the resulting slip profile via equation \eqref{eq:moment-release-definition}, we obtain the following final expression for the moment release:
\begin{equation}\label{eq:generalization-noment-volume}
    M_0(t)=\frac{16}{3(2\pi)^{3/2}} \frac{V(t)^{3/2}}{\sqrt{\Delta\tau_{r-0}}}\left( \frac{f_r}{wS} \right)^{3/2},
\end{equation}
which is identical to equation \eqref{eq:moment-volume-constant-rate} in the main text, thus demonstrating that the relation between the moment release, in-situ conditions, and injected fluid volume \eqref{eq:moment-volume-constant-rate}, holds for arbitrary fluid injections. 

Finally, we note that some constraints on the fluid source are required to ensure that some important model assumptions are satisfied. For example, the rupture must always propagate in a crack-like mode during the pressurization stage. This is necessary so that equations \eqref{eq:energy-balance} and \eqref{eq:elastic-equilibrium-circular-inverse} (in combination with \eqref{eq:stress-change}) remain always valid. Essentially, crack-like propagation allows for substituting the shear stress acting within the ruptured surface directly with the fault shear strength at any time during propagation. This would not be valid, for instance, for the pulse-like ruptures characterizing the shut-in stage \cite{Saez_Lecampion_2023}, where behind the locking front equating the shear stress to the fault strength is no longer valid. As discussed in the main text, crack-like propagation holds at least in one relevant scenario, where the pore pressure increases monotonically everywhere within the fault zone. Another assumption in the upper bound rationale of our model relies on the following property of unconditionally stable ruptures: the effect of the fracture energy in the front-localized energy balance must diminish as the rupture grows and, ultimately, become negligible \cite{Saez_Lecampion_2024}. Although this is certainly valid even in the case of arbitrary fluid injections, it relies on an implicit assumption of the slip-weakening model, namely, the fracture energy being constant. Our theoretical framework allowed in principle to account for non-constant and non-uniform fracture energy. We do not account for fracture-energy heterogeneity for the same reason that we do not account for stress or other kinds of heterogeneities in our model: we aim to provide fundamental, first-order insights into the problem at hand. Moreover, we also do not consider a possible scale dependence of the fracture energy. The scale-dependency of fracture energy for seismic ruptures is a topic of active research (\cite{Cocco_Aretusini_2023} and references therein). Although we do expect this phenomenon to be also present in aseismic ruptures, to the best of our knowledge, there is no experimental or observational evidence suggesting such behavior for slow frictional ruptures. We therefore refrain from exploring the theoretical implications of this hypothetical physical ingredient at the moment.

\subsection{Numerical methods}

All the numerical calculations in this study have been conducted via the 
boundary-element-based method described in \cite{Saez_Lecampion_2022}. For the general case of non-circular ruptures, we use the fully three-dimensional method presented in \cite{Saez_Lecampion_2022}. For the particular case of axisymmetric, circular ruptures, we use a more efficient axisymmetric version of the method presented in \cite{Saez_Lecampion_2023}.

\subsection*{Acknowledgements}
The authors thank J.-P. Avouac for providing comments on an earlier manuscript version.

 \subsection*{Funding}
The results were obtained within the EMOD project (Engineering model for hydraulic stimulation). The EMOD project benefits from a grant (research contract no. SI/502081-01) and an exploration subsidy (contract no. MF-021-GEO-ERK) of the Swiss federal office of energy for the EGS geothermal project in Haute-Sorne, canton of Jura, which is gratefully acknowledged. A.S. was partially funded by the Federal Commission for Scholarships for Foreign Students via the Swiss Government Excellence Scholarship. F.P. acknowledges funding from the European Union (ERC Starting Grant HOPE num. 101041966).

\subsection*{Author contributions}

A.S. developed the main ideas of the upper-bound model, conducted the theoretical derivations and numerical calculations, compiled and produced the dataset of slow slip events, interpreted the results, carried out the discussion, and wrote the manuscript. F.P. provided the dataset of cm-scale laboratory events, contributed to the interpretation of the laboratory data, and provided comments on the manuscript. B.L. helped to develop some of the main ideas and provided comments on the manuscript.

\subsection*{Competing interests}
The authors declare that they have no competing interests.

\printbibliography

\end{document}